\documentclass{article}
\usepackage{epsfig}
\usepackage{color}

\makeatletter
\@addtoreset{equation}{section}

\makeatother

\begin{document}

\begin{flushright}
Dec 2009

SNU09-016
\end{flushright}

\begin{center}

\vspace{5cm}

{\LARGE 
\begin{center}
On Large $N$ Solution of ABJM Theory
\end{center}
}

\vspace{2cm}

Takao Suyama \footnote{e-mail address : suyama@phya.snu.ac.kr}

\vspace{1cm}

{\it Center for Theoretical Physics, 

Seoul National University, 

Seoul 151-747, Republic of Korea}

\vspace{3cm}

{\bf Abstract} 

\end{center}

We investigate the large $N$ limit of the expectation value $W(\lambda)$ of a BPS Wilson loop in ABJM theory, using an integral 
expression of the partition function obtained recently by Kapustin et al. 
Certain saddle-point equations provide the correct perturbative expansion of $W(\lambda)$. 
The large $\lambda$ behavior of $W(\lambda)$ is also obtained from the saddle-point equations. 
The result is consistent with AdS/CFT correspondence. 

\vspace{1cm}


\newpage

\vspace{1cm}

\section{Introduction}

\vspace{5mm}

Our knowledge on M-theory has become deeper since the discovery of the worldvolume theory on M2-branes 
\cite{Bagger:2006sk}\cite{Bagger:2007jr}\cite{Bagger:2007vi}\cite{Gustavsson:2008dy}\cite{Gustavsson:2007vu}\cite{Aharony:2008ug}. 
The theory turned out to be a superconformal Chern-Simons theory coupled to matters\footnote{
BLG theory was rewritten as the Chern-Simons-matter theory in \cite{VanRaamsdonk:2008ft}. 
}. 
The relevance of the Chern-Simons theory in the context of M-theory was already anticipated in \cite{Schwarz:2004yj}. 

It is well known that a gauge theory may be simplified by considering the planar limit. 
A set of M2-branes in ${\bf C}^4/{\bf Z}_k$ background is described by ABJM theory \cite{Aharony:2008ug} 
with gauge group U$(N)\times$U$(N)$ and 
the Chern-Simons level $(k,-k)$. 
In the large $N$ limit with the 't Hoof coupling $\lambda=N/k$ kept finite, ABJM theory is expected to be dual to Type IIA 
string theory in AdS$_4\times {\bf CP}^3$ with fluxes \cite{Aharony:2008ug}. 
A natural expectation is that this correspondence might be directly checked by solving ABJM theory in the large $N$ limit 
while keeping $\lambda$ finite. 
Clearly, this must be a difficult problem. 
Instead, it may be easier to calculate a particular observable in ABJM theory in the limit, and to compare the result 
with the corresponding 
observable in Type IIA string theory. 

In AdS$_5$/CFT$_4$ correspondence, it is already known that there exists such an observable for which the above-mentioned line 
of research is possible. 
In \cite{Pestun:2007rz}, the expectation value of the half-BPS Wilson loop was exactly calculated. 
In fact, it is the exact result both for finite $N$ and finite $g_{YM}^2N$. 
It turned out that the large $N$ limit of the exact result on the Wilson loop reproduces the corresponding quantity in Type IIA 
string theory which was proposed in \cite{Rey:1998ik}\cite{Rey:1998bq}\cite{Maldacena:1998im} 
and to which strong evidence was given in \cite{Erickson:2000af}. 

After the discovery of ABJM theory, a BPS Wilson loop operator was constructed in 
\cite{Drukker:2008zx}\cite{Chen:2008bp}\cite{Rey:2008bh}, and the perturbative calculation of 
the expectation value $W(\lambda)$ was performed. 
Quite differently from the case of ${\cal N}=4$ super Yang-Mills theory, the BPS Wilson loop preserves {\it at most} $\frac16$ of 
supersymmetry. 
This fact makes it difficult to identify what should be the corresponding object to the Wilson loop in Type IIA string theory. 
However, it seems to be still reasonable to expect that the BPS Wilson loop would have a dual string worldsheet in 
AdS$_4\times{\bf CP}^3$. 
To confirm this conjecture, it is necessary to determine the large $\lambda$ behavior of $W(\lambda)$. 

Recently, a localization technique was applied to the partition function and $W(\lambda)$ of ABJM theory in 
\cite{Kapustin:2009kz}, and an integral representations of 
them were obtained. 
It was also shown that the integral representation of $W(\lambda)$ reproduces the perturbative expansion of $W(\lambda)$ 
obtained in \cite{Drukker:2008zx}\cite{Chen:2008bp}\cite{Rey:2008bh} by calculating Feynman diagrams. 

The integral representation of the partition function of ABJM theory obtained in \cite{Kapustin:2009kz} 
looks similar to the partition 
function of a matrix model where the angular variables are integrated out. 
It is natural to expect that the techniques developed for solving matrix models may be applicable to ABJM theory. 
In this paper, we will show that both the perturbative expansion of $W(\lambda)$ and the large $\lambda$ asymptotic behavior of 
$W(\lambda)$ are derived from two saddle-point equations which are obtained from the integral representation. 
The perturbative expansion of $W(\lambda)$ derived from the saddle-point equations exactly coincides with the one in 
\cite{Kapustin:2009kz}, 
including the phase factor due to the framing. 
The large $\lambda$ behavior of $W(\lambda)$ turns out to be 
\begin{equation}
W(\lambda) \sim e^{c\sqrt{\lambda}}
\end{equation}
for a constant $c>0$ whose upper bound is obtained\footnote{
In some earlier versions of the paper, we claimed that the value (\ref{coefficient}) of $c$ would be the exact value. 
After the appearance of \cite{Marino:2009jd}, the arguments were re-examined and then 
it turned out that this claim was too strong which could not be justified by the analysis 
done in this paper. 
As explained in section \ref{largelambda}, the best we can find is the upper bound on $c$ which is compatible with 
\cite{Marino:2009jd}. 
}. 
This result is consistent with the conjecture claiming that a string worldsheet in the bulk would be dual to the 
BPS Wilson loop. 
The large $N$ solution for finite $\lambda$, however, seems quite difficult to obtain, and it is still an open issue. 

This paper is organized as follows. 
The localization calculation of \cite{Kapustin:2009kz} is briefly reviewed in section \ref{localization}. 
The saddle-point equations are derived in section \ref{saddle}. 
In section \ref{perturbative}, the perturbative expansion of $W(\lambda)$, up to $O(\lambda^3)$, 
is derived from the saddle-point equations.  
Section \ref{algorithm} shows a recursive algorithm for calculating $W(\lambda)$ perturbatively. 
Based on some observations on the saddle-point equations for finite $N$ obtained in section \ref{finiteN}, we analyze the 
large $\lambda$ behavior of $W(\lambda)$ in section \ref{largelambda}. 
Section \ref{discuss} is devoted to discussion. 
Appendix \ref{higher_order} contains the details of the calculation of $W(\lambda)$ up to order $\lambda^{11}$.

\vspace{1cm}

\section{Localization for ABJM theory} \label{localization}

\vspace{5mm}

In suitable situations, the localization is a very powerful tool to 
{\it exactly} calculate some quantities of a supersymmetric theory. 
It was applied to the expectation value of the half-BPS Wilson loop in ${\cal N}=4$ super Yang-Mills theory in 
\cite{Pestun:2007rz}. 
The result is remarkably simple, namely, the path-integral for the Wilson loop reduces to an integral for the Gaussian matrix 
model which can be performed easily. 
Interestingly enough, this exact result enables us to find the behavior of the Wilson loop in the large 't Hooft coupling 
limit which coincides with the one expected from AdS/CFT correspondence. 
See also \cite{Pestun:2009nn}\cite{Giombi:2009ds}\cite{Giombi:2009ek}\cite{strings2009} for applications to other theories. 

Recently, a similar localization technique was applied to supersymmetric Chern-Simons-matter theories including ABJM theory in 
\cite{Kapustin:2009kz}. 
There is a technical simplification when the three-dimensional theories are considered. 
One can use superfields which are obtained by the dimensional reduction from ${\cal N}=1$ superfields in four-dimensions. 
In ${\cal N}=2$ gauge theories in three-dimensions, there exists a Wilson loop \cite{Gaiotto:2007qi} 
which preserves a fraction of the supersymmetry realized on the superfields {\it off-shell}. 
In the case of ${\cal N}=4$ super Yang-Mills theory, the supersymmetry which is preserved by the half-BPS Wilson loop is 
always realized only on-shell, and therefore, a sophisticated construction of an alternative off-shell symmetry was necessary 
in \cite{Pestun:2007rz}. 

ABJM theory has the gauge group U$(N)\times $U$(N)$. 
According to this, there are two kinds of Wilson loops, each of which is constructed from either a gauge field 
$A_\mu$ or $\tilde{A}_\mu$ of the two U$(N)$ factors. 
In the following, we only consider the Wilson loop for $A_\mu$. 
A BPS Wilson loop was constructed and investigated in \cite{Drukker:2008zx}\cite{Chen:2008bp}\cite{Rey:2008bh}. 
For the fundamental representation, the explicit form is 
\begin{equation}
W_{\rm N}(C) 
= \frac1N\mbox{Tr}_{\rm N}\,\mbox{P}\exp \int_Cds \Bigl[ i\dot{x}^\mu A_{\mu}(x)+|\dot{x}|M_A{}^BX^A(x)X_B(x) \Bigr].
   \label{BPSWL}
\end{equation}
Here $X^A$ are scalar fields in the bi-fundamental representation of U$(N)\times$U$(N)$, and in ${\bf 4}$ of SU$(4)$ R-symmetry. 
$M_A{}^B$ is a constant matrix. 
It turns out that a fraction of supersymmetry is preserved iff 
\begin{equation}
M_A{}^B = \mbox{diag}(+1,+1,-1,-1)
\end{equation}
up to an R-symmetry transformation. 

The localization performed in \cite{Kapustin:2009kz} goes roughly as follows. 
ABJM theory is defined on $S^3$, and the contour $C$ of the Wilson loop is placed on the equator of $S^3$. 
Let $Q$ be a fermionic transformation induced by a suitably chosen supercharge which is preserved by the Wilson loop. 
By definition, $Q^2$ is a bosonic transformation of the theory. 
Let $V=\psi^\dag Q\psi$ where $\psi$ is the collective notation for the fermions in the theory. 
If $V$ satisfies $\int d^3x\,Q^2V=0$, then the path-integral can be modified to 
\begin{equation}
Z_{ABJM} = \int e^{-S_{ABJM}-t\int d^3x\,QV}, 
\end{equation}
without changing the value of the partition function. 
The same is true for the Wilson loop: 
\begin{equation}
\langle W_{\rm N}(C) \rangle = Z_{ABJM}^{-1}\int e^{-S_{ABJM}-t\int d^3x\,QV}W_{\rm N}(C). 
\end{equation}
Those quantities are independent of $t$. 
If $t$ is taken to be large, then, since the bosonic part of $QV$ is $|Q\psi|^2$ which is positive definite, the path-integral 
is localized to a set of field configurations for which the bosonic part of $QV$ vanishes. 
The allowed configurations turn out to be coordinate independent, and therefore, the path-integral reduces to a 
finite-dimensional matrix integral. 
There could also be the one-loop contribution and the non-perturbative contributions to the integrand of the matrix integral. 
In \cite{Kapustin:2009kz}, the former was calculated explicitly, and the latter was shown to be absent. 

As a result, the partition function of ABJM theory becomes 
\begin{equation}
Z_{ABJM} = \int\prod_{i=1}^Nd\phi_id\tilde{\phi}_i\,e^{\pi ik\sum_{i=1}^N(\phi_i^2-\tilde{\phi}_i^2)}
 \frac{\prod_{i<j}\sinh^2[\pi(\phi_i-\phi_j)]\sinh^2[\pi(\tilde{\phi}_i-\tilde{\phi}_j)]}{\prod_{ij}\cosh^2[\pi(
 \phi_i-\tilde{\phi}_j)]}, 
   \label{Kapustin}
\end{equation}
and the expectation value of the Wilson loop (\ref{BPSWL}) becomes 
\begin{eqnarray}
\langle W_{\rm N}(C) \rangle 
&=& Z_{ABJM}^{-1}\int\prod_{i=1}^Nd\phi_id\tilde{\phi}_i\,\left[ \frac1N\sum_{i=1}^Ne^{2\pi \phi_i} \right]
    e^{\pi ik\sum_{i=1}^N(\phi_i^2-\tilde{\phi}_i^2)} \nonumber \\
& & \hspace{2cm}\times 
    \frac{\prod_{i<j}\sinh^2[\pi(\phi_i-\phi_j)]\sinh^2[\pi(\tilde{\phi}_i-\tilde{\phi}_j)]}{\prod_{ij}\cosh^2[\pi(
    \phi_i-\tilde{\phi}_j)]}. 
   \label{int WL}
\end{eqnarray}
The explicit expression of $\langle W_{\rm N}(C) \rangle$ was obtained perturbatively in \cite{Kapustin:2009kz} 
using (\ref{int WL}) 
which exactly reproduces the expression obtained in \cite{Drukker:2008zx}\cite{Chen:2008bp}\cite{Rey:2008bh} 
by calculating Feynman diagrams. 
Note that only planar diagrams were taken into account in \cite{Drukker:2008zx}\cite{Chen:2008bp}\cite{Rey:2008bh}, 
while (\ref{int WL}) provides the expression exact in both 
$k$ and $N$.

\vspace{1cm}

\section{Saddle-point equations} \label{saddle}

\vspace{5mm}

In the previous section, we briefly reviewed the derivation of 
an expression (\ref{Kapustin}) of the partition function of ABJM theory in terms of a finite-dimensional 
integral. 
This is an exact formula with respect to $N$ and $k$. 
If one is only interested in the large $N$ limit, then the necessary information should be contained in the saddle-points 
of the integral. 
Indeed, it is sufficient to obtain the large $N$ results when one would like to compare some results on ABJM theory with the 
corresponding results on the classical gravity in AdS$_4\times {\bf CP}^3$. 
This saddle-point approximation would be a more efficient way to extract information on observables in ABJM theory than the exact 
integral expression, at the cost of losing 
all $1/N$ corrections. 
Those corrections would possibly be discussed starting with the saddle-point results, but it is beyond the scope of this paper. 

\vspace{5mm}

The saddle-points of the integral (\ref{Kapustin}) are obtained as the extrema of the effective action 
\begin{eqnarray}
S_{\rm eff} 
&=& -\pi ik\sum_{i=1}^N(\phi_i^2-\tilde{\phi}_i^2) \nonumber \\
& & -\sum_{i<j}\left\{ \log\sinh^2[\pi(\phi_i-\phi_j)]+\log\sinh^2[\pi(\tilde{\phi}_i-\tilde{\phi}_j)] \right\} \nonumber \\
& & +\sum_{ij}\log\cosh^2[\pi(\phi_i-\tilde{\phi}_j)]. 
\end{eqnarray}
The saddle-point equations derived from $S_{\rm eff}$ are 
\begin{eqnarray}
-ik\phi_i &=& \sum_{j\ne i}\coth[\pi(\phi_i-\phi_j)]-\sum_{j}\tanh[\pi(\phi_i-\tilde{\phi}_j)], 
   \label{saddle1} \\
ik\tilde{\phi}_i &=& \sum_{j\ne i}\coth[\pi(\tilde{\phi}_i-\tilde{\phi}_j)]-\sum_{j}\tanh[\pi(\tilde{\phi}_i-\phi_j)]. 
   \label{saddle2}
\end{eqnarray}
It is important to notice that the coefficients in the left-hand side of (\ref{saddle1})(\ref{saddle2}) are purely imaginary. 
Therefore, this set of saddle-point equations cannot have a solution in which all $\phi_i$ and $\tilde{\phi}_i$ are real, 
although it is expected from the derivation of (\ref{Kapustin}). 
In addition, one can easily show that a solution in which all $\phi_i$ and $\tilde{\phi}_i$ are purely imaginary is not allowed. 
We assume that the integration contours in (\ref{Kapustin}) can be suitably deformed away from the real axis 
so that a complex solution of the 
saddle-point equations provides the dominant contribution to the integral. 
As we will show below, the result of the 
perturbative calculation of the Wilson loop expectation value seems to support the validity of this assumption. 

It will turn out in the next section that it is convenient to rescale the variables for the perturbative calculation. 
We define $x_i$ and $\tilde{x}_i$ by 
\begin{equation}
\phi_i = \frac{\sqrt{\lambda}}\pi x_i, \hspace{5mm} \tilde{\phi}_i = \frac{\sqrt{\lambda}}\pi \tilde{x}_i,
\end{equation}
where $\lambda=\frac Nk$ is the 't Hooft coupling. 
In terms of these new variables, the saddle-point equations (\ref{saddle1})(\ref{saddle2}) become 
\begin{eqnarray}
-\frac i\pi x_i 
&=& \frac1N\sum_{j\ne i}\sqrt{\lambda}\coth[\sqrt{\lambda}(x_i-x_j)]
   -\frac1N\sum_{j}\sqrt{\lambda}\tanh[\sqrt{\lambda}(x_i-\tilde{x}_j)], 
   \label{saddle3} \\
\frac i\pi \tilde{x}_i 
&=& \frac1N\sum_{j\ne i}\sqrt{\lambda}\coth[\sqrt{\lambda}(\tilde{x}_i-\tilde{x}_j)]
   -\frac1N\sum_{j}\sqrt{\lambda}\tanh[\sqrt{\lambda}(\tilde{x}_i-x_j)]. 
   \label{saddle4}
\end{eqnarray}

\vspace{5mm}

Let us consider the large $N$ limit of (\ref{saddle3})(\ref{saddle4}). 
We would like to discuss the limit: 
\begin{equation}
N \to \infty, \hspace{5mm} k \to \infty, \hspace{5mm} \lambda = \mbox{fixed.}
\end{equation}
In this limit, the distributions of $x_i$ and $\tilde{x}_i$ are expected to become continuous. 
Let $\rho(x)$ be the distribution 
function of $x_i$ defined on a curve $I$ in ${\bf C}$, and let $\tilde{\rho}(x)$ be such a function 
for $\tilde{x}_i$ defined on $\tilde{I}$ in ${\bf C}$. 
As mentioned above, the curves $I,\tilde{I}$ do not lie on the real axis, nor on the imaginary axis. 
The functions $\rho(x)$ and $\tilde{\rho}(x)$ are required to satisfy 
\begin{equation}
\int_I dx\,\rho(x) \  =\   1  \ = \  \int_{\tilde{I}}dx\,\tilde{\rho}(x). 
\end{equation}

In the large $N$ limit, Eqs.(\ref{saddle3})(\ref{saddle4}) are written 
in terms of $\rho(x)$ and $\tilde{\rho}(x)$  as 
\begin{eqnarray}
-\frac i\pi x 
&=& \int_I\hspace{-3.75mm}-\hspace{1mm}dx'\, \rho(x')\sqrt{\lambda}\coth[\sqrt{\lambda}(x-x')]
   -\int_{\tilde{I}}dx'\, \tilde{\rho}(x')\sqrt{\lambda}\tanh[\sqrt{\lambda}(x-x')], 
   \label{saddle5} \\
\frac i\pi \tilde{x} 
&=& \int_{\tilde{I}}\hspace{-3.75mm}-\hspace{1mm}dx'\, \tilde{\rho}(x')\sqrt{\lambda}\coth[\sqrt{\lambda}(\tilde{x}-x')]
   -\int_{I}dx'\, \rho(x')\sqrt{\lambda}\tanh[\sqrt{\lambda}(\tilde{x}-x')], 
   \label{saddle6}
\end{eqnarray}
where 
$x\in I$ and $\tilde{x}\in\tilde{I}$ 
are assumed. 

The expectation value of the Wilson loop (\ref{int WL}) is 
\begin{equation}
W(\lambda) \  =\  \int_Idx\, \rho(x)e^{2\sqrt{\lambda} x}. 
   \label{WL}
\end{equation}

\vspace{5mm}

\vspace{1cm}

\section{Perturbative calculation of Wilson loop} \label{perturbative}

\vspace{5mm}

In this section, we calculate $W(\lambda)$ perturbatively in terms of small $\lambda$. 
By expanding the right-hand side of (\ref{saddle5}) with respect to $\lambda$, and truncating terms of order $\lambda^{n+1}$ or 
higher in the resulting 
power series, 
one obtains the following truncated saddle-point equation 
\begin{equation}
-\frac i\pi x = \int_{I_n}\hspace{-5.25mm}-\hspace{2mm}dx'\, \frac{\rho_n(x')}{x-x'}+\sum_{k=1}^n\lambda^k
 \Bigl[\ c_k\Bigl\langle (x-\hat{x})^{2k-1} \Bigr\rangle_n-t_k\Bigl\langle (x-\hat{\tilde{x}})^{2k-1} \Bigr\rangle_n\ \Bigr], 
   \label{order n}
\end{equation}
where 
\begin{eqnarray}
\langle (x-\hat{x})^k \rangle_n &=& \int_{I_n}dX'\, \rho_n(x')(x-x')^k, \\
\langle (x-\hat{\tilde{x}})^k \rangle_n &=& \int_{\tilde{I}_n}dx'\, \tilde{\rho}_n(x')(x-x')^k 
\end{eqnarray}
for a positive integer $k$, and the coefficients $c_k,t_k$ are defined as 
\begin{eqnarray}
\coth x &=& \frac1x+\sum_{k=1}^\infty c_kx^{2k-1}, \\
\tanh x &=& \sum_{k=1}^\infty t_kx^{2k-1}. 
\end{eqnarray}
The function $\rho_n(x)$ is related to $\rho(x)$ as 
\begin{equation}
\rho_n(x) \ =\  \rho(x) \mbox{\ \ mod }\lambda^{n+1}. 
\end{equation}
$\tilde{\rho}_n(x)$ satisfies the corresponding equation similar to (\ref{order n}). 
Note that $I_n,\tilde{I}_n$ may change by varying $n$. 

Remarkably, for the determination of $\rho_n(x)$, it is sufficient 
to evaluate $\langle (x-\hat{x})^k \rangle_n$ and $\langle (x-\hat{\tilde{x}})^k \rangle_n$ only up to order 
$\lambda^{n-1}$ since they are always multiplied by $\lambda$ in (\ref{order n}). 
Therefore, assuming that $\rho_{n-1}(x)$ and $\tilde{\rho}_{n-1}(x)$ are 
already known, the sum in (\ref{order n}) can be evaluated explicitly. 
Then, (\ref{order n}) can be rewritten in a form of the equation which 
is familiar in the context of the one-matrix model. 
(See (\ref{saddle n}) below.)  
Namely, the sum in the right-hand side, moved to the left-hand side, is regarded as a contribution from a matrix model 
action which is a polynomial of a matrix of degree $2n-1$. 
This saddle-point equation can be solved by the well-known techniques, and therefore one can determine $\rho_n(x)$. 
In this way, $\rho(x)$ can be determined perturbatively in terms of small $\lambda$. 
$\tilde{\rho}(x)$ is determined similarly. 

$W(\lambda)$ is calculated via the formula (\ref{WL}). 
The perturbative expression of $W(\lambda)$ up to order $\lambda^{n}$ is 
\begin{equation}
W(\lambda) = 1+\sum_{k=1}^{2n}\frac{(2\sqrt{\lambda})^k}{k!}\langle \hat{x}^k \rangle+O(\lambda^{n+1}), 
\end{equation}
obtained by expanding the integrand. 
Here $\langle \hat{x}^k \rangle$ are the averages calculated in terms of $\rho(x)$. 
It will be shown later that $\langle \hat{x}^k \rangle$ with $k$ odd vanishes to all orders in $\lambda$, 
implying that $W(\lambda)$ is indeed a series of $\lambda$. 
At this order of $\lambda$, $\langle \hat{x}^k \rangle$ can be replaced with $\langle \hat{x}^k \rangle_{n-1}$ 
since these are always multiplied by $\lambda$. 
Therefore, knowing $\rho_{n-1}(x)$, $W(\lambda)$ is determined up to order $\lambda^{n}$. 

In the following, we will show explicit calculations for lower orders of $\lambda$. 

\vspace{5mm}

\subsection{$O(\lambda^0)$}

\vspace{5mm}

At this order, (\ref{order n}) becomes 
\begin{equation}
-\frac i\pi x = \int_{I_0}\hspace{-5mm}-\hspace{2mm}dx'\,\frac{\rho_0(x')}{x-x'}. 
\end{equation}
Note that $\rho_0(x)$ decouples from $\tilde{\rho}_0(x)$. 
If the coefficient of the left-hand side is real and positive, then this equation can be solved easily. 
We will use the following trick. 
We first solve 
\begin{equation}
\frac 1\kappa x = \int_{I_0}\hspace{-5mm}-\hspace{2mm}dx'\,\frac{\rho_0(x')}{x-x'}, 
   \label{gaussian}
\end{equation}
assuming $\kappa$ being real and positive, and then set $\kappa=\pi i$. 
Let us define the resolvent 
\begin{equation}
R_n(x) := \int_{I_n}dx'\,\frac{\rho_n(x')}{x-x'} \hspace{5mm} (x\notin I_n) 
\end{equation}
for each order of $\lambda$. 
$R_0(x)$ has the form 
\begin{eqnarray}
R_0(x) 
&=& \frac1\kappa x-\frac1\kappa\sqrt{x^2-2\kappa} \nonumber \\
&=& \frac1x+\frac {\kappa}{2x^3}+\frac{\kappa^2}{2x^5}+\cdots. 
   \label{0th}
\end{eqnarray}
It is important to note that $\langle \hat{x}^{k} \rangle_0=0$ for odd $k$. 

The function $\tilde{\rho}_0(x)$ satisfies (\ref{gaussian}) with $\kappa=-\pi i$. 
Therefore, all the properties of $\tilde{R}_0(x)$, defined similarly to $R_0(x)$, 
are derived easily from those of $R_0(x)$. 

The averages $\langle \hat{x}^k \rangle_n$ are determined by $R_n(x)$ as 
\begin{equation}
R_n(x) = \frac1x+\sum_{k=1}^\infty \langle \hat{x}^k \rangle_nx^{-k-1}.
   \label{<x^k>}
\end{equation}
Using the expansion (\ref{0th}), $W(\lambda)$ can be determined up to order $\lambda$, as mentioned above. 
We obtain 
\begin{equation}
W(\lambda) = 1+\pi i\lambda +O(\lambda^2). 
\end{equation}
Indeed, this result agrees with that of \cite{Kapustin:2009kz}. 

\vspace{5mm}

\subsection{$O(\lambda^1)$}

\vspace{5mm}

The saddle-point equation (\ref{order n}) becomes 
\begin{equation}
-\frac i\pi x = \int_{I_1}\hspace{-5mm}-\hspace{2mm}dx'\,\frac{\rho_1(x')}{x-x'}+\lambda\left[ \frac13\langle x-\hat{x} \rangle_0 
 -\langle x-\hat{\tilde{x}} \rangle_0 \right]. 
   \label{order 1}
\end{equation}
Using the results $\langle \hat{x} \rangle_0=\langle \hat{\tilde{x}} \rangle_0=0$ obtained in the previous subsection, 
(\ref{order 1}) can be written as 
\begin{equation}
\left[ \frac23\lambda-\frac i\pi \right]x = \int_{I_1}\hspace{-5mm}-\hspace{2mm}dx'\,\frac{\rho_1(x')}{x-x'}. 
\end{equation}
This is again the equation of the form (\ref{gaussian}) with 
\begin{equation}
\kappa^{-1} = \frac23\lambda-\frac i\pi. 
   \label{1st}
\end{equation}
As a result, $R_1(x)$ has the form (\ref{0th}). 
$W(\lambda)$ is determined up to order $\lambda^2$, and the resulting expression is 
\begin{equation}
W(\lambda) = 1+\pi i\lambda+\frac13\pi^2\lambda^2+O(\lambda^3). 
\end{equation}
This coincides exactly with the result obtained in \cite{Kapustin:2009kz}. 
To see the agreement with the perturbative calculation \cite{Drukker:2008zx}\cite{Chen:2008bp}\cite{Rey:2008bh}, 
it is convenient to pull out the phase factor $e^{\pi i\lambda}$ due to the framing \cite{Kapustin:2009kz}. 
The result is 
\begin{equation}
W(\lambda) = e^{\pi i\lambda}\left[ 1+\frac56\pi^2\lambda^2+O(\lambda^3) \right]. 
\end{equation}

\vspace{5mm}

\subsection{$O(\lambda^2)$}

\vspace{5mm}

At this order, a generic phenomenon occurs in the perturbative calculation. 
Namely, the saddle-point equation (\ref{order n}) 
we have to solve becomes more complicated than the Gaussian one (\ref{gaussian}). 

Eq.(\ref{order n}) at this order becomes 
\begin{equation}
-\frac i\pi x = \int_{I_2}\hspace{-5mm}-\hspace{2mm}dx'\,\frac{\rho_2(x')}{x-x'}+\lambda\left[ \frac13\langle x-\hat{x} 
 \rangle_1 
 -\langle x-\hat{\tilde{x}} \rangle_1 \right]+\lambda^2\left[ -\frac1{45}\langle (x-\hat{x})^3 \rangle_0
 +\frac13\langle (x-\hat{\tilde{x}})^3 
 \rangle_0 \right]. 
\end{equation}
Using the result (\ref{0th}) with (\ref{1st}), this equation can be written as 
\begin{equation}
\alpha_2x^3+\alpha_1x = \int_{I_2}\hspace{-5mm}-\hspace{2mm}dx'\,\frac{\rho_2(x')}{x-x'}, 
\end{equation}
where 
\begin{eqnarray}
\alpha_2 &=& -\frac{14}{45}\lambda^2, \\
\alpha_1 &=& -\frac i\pi+\frac23\lambda+\frac{8\pi i}{15}\lambda^2. 
\end{eqnarray}
This saddle-point equation can be solved by the well-known technique. 
The resolvent should have the form 
\begin{equation}
R_2(x) = \alpha_2x^3+\alpha_1x-(\beta_2x^2+\beta_1)\sqrt{x^2-\gamma}. 
\end{equation}
The constants $\beta_2,\beta_1$ and $\gamma$ are determined by requiring 
\begin{equation}
\lim_{x\to\infty}xR_2(x) = 1. 
\end{equation}
The result is 
\begin{eqnarray}
\beta_2 &=& -\frac{14}{45}\lambda^2, \\
\beta_1 &=& -\frac i\pi+\frac23\lambda+\frac{2\pi i}{9}\lambda^2 \\
\gamma &=& 2\pi i+\frac{4\pi^2}3\lambda-\frac{34\pi^3 i}{45}\lambda^2. 
\end{eqnarray}
The expansion of $R_2(x)$ in terms of $x^{-1}$ provides 
\begin{eqnarray}
\langle \hat{x}^2 \rangle_2 &=& \frac{\pi i}2+\frac{\pi^2}3\lambda-\frac{\pi^3i}9\lambda^2, \\
\langle \hat{x}^4 \rangle_2 &=& -\frac{\pi^2}2+\frac{2\pi^3 i}3\lambda+\frac{29\pi^4}{60}\lambda^2, \\
\langle \hat{x}^6 \rangle_2 &=& -\frac{5\pi^3i}8-\frac{5\pi^4}4\lambda+\frac{41\pi^5i}{30}\lambda^2, 
\end{eqnarray}
which are used to determine $W(\lambda)$ as follows, 
\begin{eqnarray}
W(\lambda) 
&=& 1+i\pi \lambda+\frac13\pi^2\lambda+\frac{i}6\pi^3\lambda^3+O(\lambda^4) \nonumber \\
&=& e^{\pi i\lambda}\left[ 1+\frac56\pi^2\lambda^2-\frac i2\pi^3\lambda^3 +O(\lambda^4) \right].
   \label{3rd}
\end{eqnarray}
We obtained the perturbative expression which agrees exactly with \cite{Kapustin:2009kz}\footnote{
An argument was given in \cite{Rey:2008bh} showing that $W(\lambda)$ should be a series of $\lambda^2$. 
The argument was based on the use of a regularization of the Wilson loop in which the loop lies on a two-plane. 
This would be realized by introducing a set of concentric loops with slightly different radii, but a regularization like this 
would break supersymmetry. 
As pointed out in \cite{Kapustin:2009kz}, in our calculation, 
a non-trivial framing would be introduced as a regularization in which the loops cannot lie 
on a two-plane for preserving supersymmetry. 
As a result, the argument in \cite{Rey:2008bh} 
cannot apply to our calculation, allowing the appearance of the $\lambda^3$ term in 
(\ref{3rd}). 
}.

\vspace{1cm}

\section{A recursive algorithm} \label{algorithm}

\vspace{5mm}

The calculation of $W(\lambda)$ shown above can be done systematically. 
In fact, there exists a recursive algorithm to determine the resolvent $R_n(x), \tilde{R}_n(x)$ from 
$R_{n-1}(x), \tilde{R}_{n-1}(x)$. 
In this section, we show this algorithm. 
We start with the suitable choice of an ansatz for the resolvents at each order of $\lambda$. 

\vspace{5mm}

\subsection{Ansatz for the resolvent}

\vspace{5mm}

We claim that, for every positive integer $n$, the resolvent $R_n(x)$ and $\tilde{R}_n(x)$ should have the form 
\begin{eqnarray}
R_n(x) 
&=& \sum_{k=1}^n\alpha_kx^{2k-1}-\left( \sum_{k=1}^n\beta_kx^{2k-2} \right)\sqrt{x^2-\gamma}, 
    \label{ansatz} \\
\tilde{R}_n(x) 
&=& \sum_{k=1}^n\tilde{\alpha}_kx^{2k-1}-\left( \sum_{k=1}^n\tilde{\beta}_kx^{2k-2} \right)\sqrt{x^2-\tilde{\gamma}}, 
    \label{ansatz2}
\end{eqnarray}
where $\alpha_k$, $\beta_k$ etc. are polynomials of $\lambda$ of degree $n$. 
The orders, i.e. the minimum power of $\lambda$ with non-zero coefficient, 
of $\alpha_k$ and $\beta_k$ with $k>1$ are $k$, while those of $\alpha_1, \beta_1$ and $c$ are 0. 
$\tilde{\alpha}_k$ etc. have the similar properties. 

We have shown that $R_1(x)$ has the above form, and it is easy to show that $\tilde{R}_1(x)$ can be obtained from $R_1(x)$ by 
\begin{equation}
\tilde{\alpha}_k = \alpha_k^*, \hspace{5mm} \tilde{\beta}_k = \beta_k^*, \hspace{5mm} \tilde{\gamma} = \gamma^*. 
   \label{cc}
\end{equation}
Suppose that $R_{n-1}(x), \tilde{R}_{n-1}(x)$ have the claimed form. 
These forms of the resolvents imply 
\begin{equation}
\langle \hat{x}^k \rangle_{n-1} = 0, \hspace{5mm} \langle \hat{\tilde{x}}^k \rangle_{n-1} = 0. 
   \label{vanish}
\end{equation}
for odd $k$. 

Let us consider the saddle-point equations at order $\lambda^n$. 
(\ref{order n}) can be written as 
\begin{equation}
\sum_{k=1}^n\alpha_kx^{2k-1} = \int_{I_n}\hspace{-5.25mm}-\hspace{2mm}dx'\,\frac{\rho_n(x')}{x-x'}, 
   \label{saddle n}
\end{equation}
where 
\begin{equation}
\sum_{k=1}^n\alpha_kx^{2k-1} 
 = -\frac i\pi x-\sum_{k=1}^{n}c_k\lambda^{k}\Bigl\langle (x-\hat{x})^{2k-1} \Bigr\rangle_{n-1}
    +\sum_{k=1}^{n}t_k\lambda^{k}\Bigl\langle (x-\hat{\tilde{x}})^{2k-1} \Bigr\rangle_{n-1}. 
   \label{force}
\end{equation}

The right-hand side of (\ref{force}) 
is indeed an odd polynomial of $x$ of degree $2n-1$ since a term with an even power of $x$ is multiplied 
by $\langle \hat{x}^k \rangle_{n-1}$ or $\langle \hat{\tilde{x}}^k \rangle_{n-1}$ 
with $k$ odd which vanish as shown in (\ref{vanish}). 
As a result, the distribution function $\rho_n(x)$ is also symmetric at order $\lambda^n$. 
For this saddle-point equation, the ansatz (\ref{ansatz}) is the suitable choice\footnote{
This is the suitable choice as long as we restrict ourselves to a one-cut solution. 
Although the right-hand side of (\ref{force}) is a polynomial of a high degree in general, those would not change the structure 
of the cut drastically since each monomial of a high degree is always multiplied by a high power of $\lambda$, and therefore 
the effect should not be relevant.
}. 
By induction, the resolvent $R_n(x)$ has the form (\ref{ansatz}) to all orders in $\lambda$. 
The ansatz (\ref{ansatz2}) for $\tilde{R}_n(x)$ is also deduced similarly. 
As a corollary, the eigenvalue distributions are symmetric to all orders in $\lambda$, implying 
$\langle \hat{x}^k \rangle=\langle \hat{\tilde{x}}^k \rangle=0$ for odd $k$. 

The coefficients $\alpha_k$ can be written explicitly as 
\begin{equation}
\alpha_k = \sum_{l=k}^n\lambda^l{{2l-1} \choose {2k-1}}\left[ -c_l\Bigl\langle \hat{x}^{2(l-k)} \Bigr\rangle_{n-1}
 +t_l\Bigl\langle \hat{\tilde{x}}^{2(l-k)} \Bigr\rangle_{n-1} \right]-\frac i\pi\delta_{k,1}. 
   \label{alpha_k}
\end{equation}
This can be calculated using $R_{n-1}(x)$ and $\tilde{R}_{n-1}(x)$. 

Since the saddle-point equations for $\rho_n(x)$ and $\tilde{\rho}_n(x)$ decouple, they can be solved separately. 
In the following, we focus on the solution for $R_n(x)$. 
$\tilde{R}_n(x)$ can be determined similarly. 

\vspace{5mm}

\subsection{Determination of $R_n(x)$}

\vspace{5mm}

The resolvent $R_n(x)$ is assumed to be of the form (\ref{ansatz}) in which $\alpha_k$ are given as (\ref{alpha_k}). 
Then, $\beta_k$ and $\gamma$ are determined by the requirement 
\begin{equation}
\lim_{x\to\infty}xR_n(x) = 1. 
   \label{condition}
\end{equation}
This is equivalent to the following requirement 
\begin{equation}
\lim_{x\to\infty}\frac{x^2}{\sqrt{x^2-\gamma}}R_n(x) = 1, 
   \label{condition2}
\end{equation}
which turns out to be more convenient. 

The Taylor expansion of $(x^2-\gamma)^{-\frac12}R_n(x)$ is 
\begin{equation}
\begin{array}{rrrcrrr}
-\beta_nx^{2n-2} & -\beta_{n-1}x^{2n-4} & -\beta_{n-2}x^{2n-6} & \cdots & -\beta_1 &  & \\
+\alpha_nx^{2n-2} & +p_1\gamma\alpha_{n}x^{2n-4} & +p_2\gamma^2\alpha_{n}x^{2n-6} & \cdots & +p_{n-1}\gamma^{n-1}\alpha_{n} & 
 +p_{n}\gamma^{n}\alpha_{n}x^{-2} & 
 \cdots \\
 & +\alpha_{n-1}x^{2n-4} & +p_1\gamma\alpha_{n-1}x^{2n-6} & \cdots & +p_{n-2}\gamma^{n-2}\alpha_{n-1} & 
 +p_{n-1}\gamma^{n-1}\alpha_{n-1}x^{-2} 
 & \cdots \\
 & & +\alpha_{n-2}x^{2n-6} & \cdots & +p_{n-3}\gamma^{n-3}\alpha_{n-2} & +p_{n-2}\gamma^{n-2}\alpha_{n-2}x^{-2} & \cdots \\
 & & & & \vdots & \vdots & \\
 & & & & +\alpha_1 & +p_1\gamma\alpha_1x^{-2} & \cdots 
\end{array}
\end{equation}
where $p_n$ are defined as 
\begin{equation}
(1-x)^{-\frac12} = \sum_{k=0}^\infty p_kx^k, \hspace{5mm} p_k = \frac1{\sqrt{\pi}}\frac{\Gamma(k+\frac12)}{k!}. 
\end{equation}
The condition (\ref{condition2}) determines $\beta_k$ in terms of $\alpha_k$ and $\gamma$ as 
\begin{equation}
\beta_k = \sum_{i=k}^np_{i-k}\gamma^{i-k}\alpha_i. 
   \label{beta_k}
\end{equation}
Let $\gamma$ be 
\begin{equation}
\gamma = \sum_{k=0}^{n}\gamma_k\lambda^k + O(\lambda^{n+1}). 
\end{equation}
$\gamma_k$ with $k<n$ are obtained from $R_{n-1}(x)$, and therefore, they are supposed to be known. 
Since the order of $\alpha_k$ with $k>1$ is larger than 1, 
(\ref{beta_k}) determines $\beta_k$ up to order $\lambda^n$ without knowing $\gamma_n$. 
The remaining unknown constant $\gamma_n$ is then determined by requiring 
\begin{equation}
\sum_{i=1}^np_i\gamma^i\alpha_i = 1. 
\end{equation}
This is equivalent to a linear equation for $\gamma_n$ which can be solved easily. 

This completes the determination of $R_n(x)$. 
By induction, it can be shown that $\tilde{R}_n(x)$ is obtained from $R_n(x)$ by (\ref{cc}). 

The expansion of $R_n(x)$ in terms of $x^{-1}$ provides $\langle \hat{x}^k \rangle_n$ via (\ref{<x^k>}). 
The explicit form of them is  
\begin{equation}
\langle \hat{x}^{2k} \rangle_n = -\sum_{i=1}^n q_{i+k}\gamma^{i+k}\beta_i,  
\end{equation}
where 
\begin{equation}
(1-x)^{\frac12} = \sum_{k=0}^\infty q_kx^k, \hspace{5mm} q_k = -\frac1{2\sqrt{\pi}}\frac{\Gamma(k-\frac12)}{k!}. 
\end{equation}
These quantities are used to determine $\alpha_k$ at the next order by (\ref{alpha_k}). 
In this way, $R_n(x)$ can be determined recursively. 

\vspace{5mm}

Recall that the vev of the Wilson loop $W(\lambda)$ is obtained as 
\begin{equation}
W(\lambda) = 1+\sum_{k=1}^{n+1}\frac{(4\lambda)^k}{(2k)!}\langle x^{2k} \rangle_n + O(\lambda^{n+2}). 
\end{equation}

We performed the calculation of $W(\lambda)$ up to order $\lambda^{11}$, according to the algorithm described above. 
The result of $W(\lambda)$ is\footnote{
In some earlier versions of the paper, the coefficient of $\lambda^{11}$ was not correct. 
After the appearance of \cite{Marino:2009jd}, the calculations were re-examined, and then a small mistake was found 
which affected only the 
coefficient of $\lambda^{11}$. 
By fixing it, we found that the result perfectly coincided with that in \cite{Marino:2009jd} up to the order we calculated. 
}
\begin{eqnarray}
W(\lambda) 
&=& e^{\pi i\lambda}\left[ 1+{{5\pi^2\lambda^2}\over{6}}-{{i\pi^3\lambda^3}\over{2}}-{{29\pi^4\lambda^4 }\over{120}}
    +{{i\pi^5\lambda^5}\over{12}}
    +{{151\pi^6\lambda^6}\over{1008}}-{{i\pi^7\lambda^7}\over{10}}-{{87449\pi^8\lambda^8}\over{362880}} \right. \nonumber \\
& & \left. +{{2603i\pi^9\lambda^9}\over{15120 }}+{{3447391\pi^{10}\lambda^{10}}\over{7983360}}
    -{{1166161i\pi^{11}\lambda^{11}}\over{3628800}} +O(\lambda^{12}) \right]. 
\end{eqnarray}
Further details of the result is summarized in appendix \ref{higher_order}.

\vspace{1cm}

\section{Solutions of the saddle-point equations for finite $N$} \label{finiteN}

\vspace{5mm}

We have shown that the saddle-point equations (\ref{saddle1})(\ref{saddle2}) 
provide the correct perturbative expansion of the Wilson loop $W(\lambda)$. 
This shows that the saddle-point equations are convenient tools to provide perturbative results of ABJM theory, enabling 
one to bypass complicated calculations of Feynman diagrams. 
Using the algorithm shown in the previous section, one can systematically calculate $W(\lambda)$ up to any desired order. 
Other observables in ABJM theory may also be calculable provided that they are given in terms of $\phi_i$ and $\tilde{\phi}_i$. 

However, this perturbative calculation, although it makes the calculations much easier, does not provide information on 
the large $\lambda$ behavior of $W(\lambda)$ which is one of the central issues in AdS/CFT correspondence. 
The correspondence between Wilson loops and string worldsheets has been well-established in AdS$_5$/CFT$_4$ case 
\cite{Rey:1998ik}\cite{Rey:1998bq}\cite{Maldacena:1998im}, and 
it would be natural to expect that a similar correspondence holds in AdS$_4$/CFT$_3$ case 
\cite{Drukker:2008zx}\cite{Chen:2008bp}\cite{Rey:2008bh}. 
This conjecture would imply that $W(\lambda)$ would behave as 
\begin{equation}
W(\lambda) \sim e^{c\sqrt{\lambda}}
   \label{conjecture}
\end{equation}
with a suitable constant $c$ for large $\lambda$. 
This result is expected to hold in the large $N$ limit with $\lambda$ large but finite. 
Therefore, it is natural to expect that the saddle-point equations might be useful to check the validity of the conjecture. 

In this section, we solve (\ref{saddle1})(\ref{saddle2}) for $N=3$ and arbitrary $k$ numerically. 
Of course, this investigation cannot give any definite results on the large $\lambda$ behavior of $W(\lambda)$. 
However, we will see that the results for $N=3$ may provide us a rough idea on the dependence of the distributions of 
$\phi_i,\tilde{\phi}_i$ on 
the value of $k$, that is, on $\lambda$. 
In the next section, we will discuss an ansatz for the distributions for large $\lambda$ 
based on the observation in this section. 
We will argue that the ansatz seems to be compatible with the behavior (\ref{conjecture}). 

\vspace{5mm}

One may notice that (\ref{saddle1})(\ref{saddle2}) have symmetries. 
These are invariant under the simultaneous sign flips $\{\phi_i,\tilde{\phi}_i\}\to \{-\phi_i,-\tilde{\phi}_i\}$. 
In addition, (\ref{saddle2}) is obtained by taking the complex conjugation of (\ref{saddle1}) accompanied by a 
replacement $\phi_i^*\leftrightarrow \tilde{\phi}_i$. 
It is clear from these symmetries that, provided that $\{\phi_i,\tilde{\phi}_i\}$ is a solution of 
(\ref{saddle1})(\ref{saddle2}), 
then 
\begin{equation}
\{-\phi_i,-\tilde{\phi}_i\}, \hspace{5mm} \{\tilde{\phi}^*_i,\phi^*_i\}, \hspace{5mm} 
 \{-\tilde{\phi}^*_i,-\phi^*_i\}, \hspace{5mm} 
\end{equation}
are also solutions. 
We are led to assume that there is a solution of the form $\{\phi_i,\phi_i^*\}$. 
Under this assumption, 
(\ref{saddle1}) and (\ref{saddle2}) are equivalent. 
Therefore, it is enough to solve (\ref{saddle1}) only. 

\begin{figure}
\epsfxsize=2.2in
\begin{center}
\includegraphics[scale=0.5]{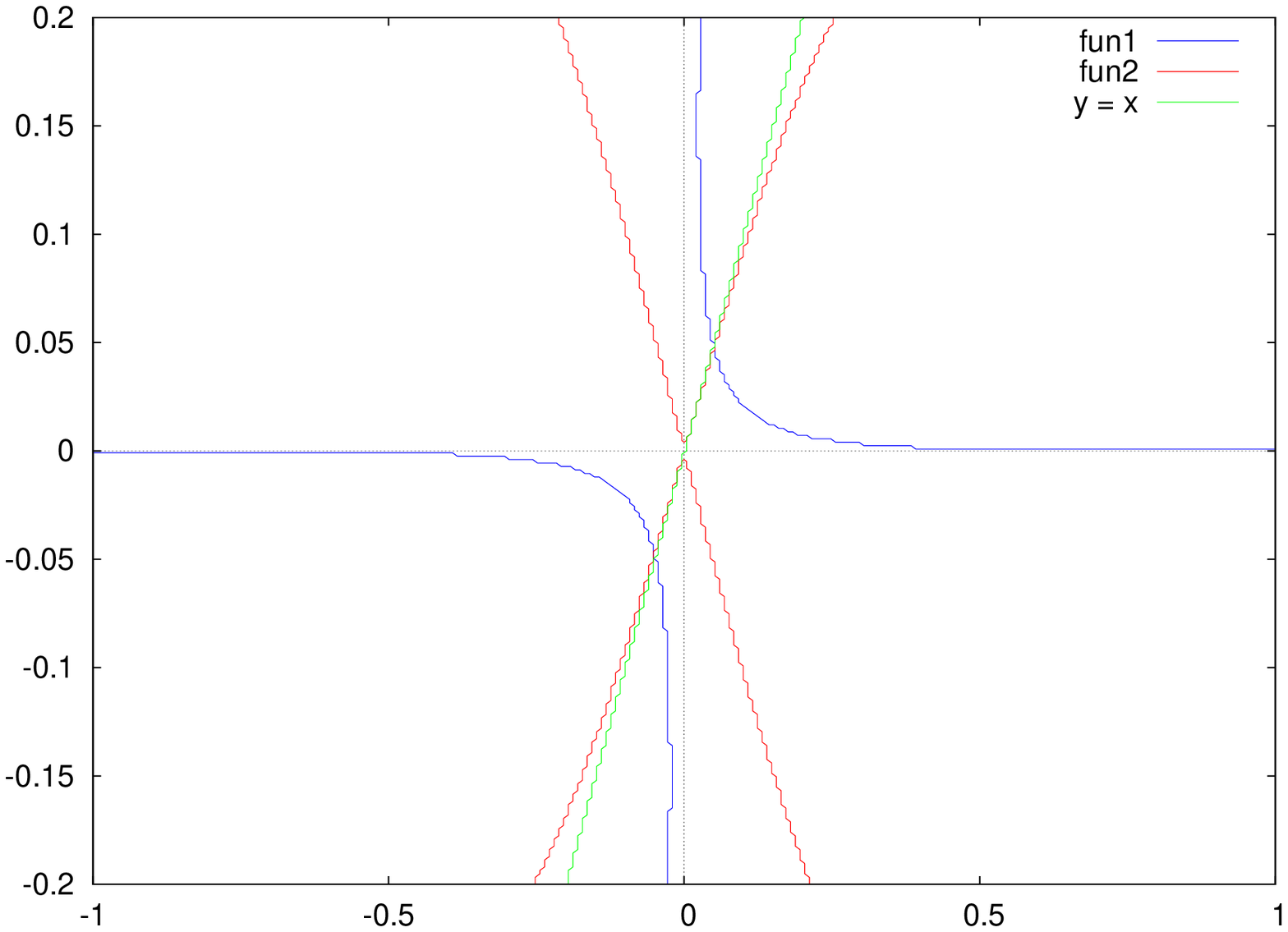}
\end{center}
\caption{Solution of the saddle-point equations for $k=100$. }
\label{k=100}
\end{figure}

\begin{figure}
\epsfxsize=2.2in
\begin{center}
\includegraphics[scale=0.5]{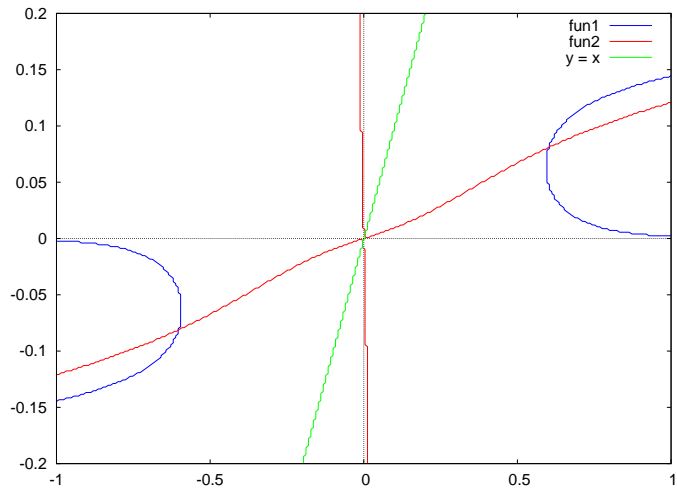}
\end{center}
\caption{Solution of the saddle-point equations for $k=1$. }
\label{k=1}
\end{figure}

Now let us focus on the case $N=3$. 
By the symmetries mentioned above, it is natural to assume 
\begin{equation}
\phi_1 = z, \hspace{5mm} \phi_2 = 0, \hspace{5mm} \phi_3 = -z. 
\end{equation}
The only non-trivial equation is now 
\begin{equation}
F(z) := ikz + \coth(\pi z)+\coth(2\pi z)-\tanh(2\pi{\rm Re}(z))-\tan(2\pi{\rm Im}(z))-\tanh(\pi z) = 0. 
\end{equation}
The graphs for Re$F(z)=0$ (fun1) and Im$F(z)=0$ (fun2) are plotted in Fig.\ref{k=100} and Fig.\ref{k=1}. 
Here we focus on the solution which is the closest to $z=0$. 

Note that the solution is almost on the line $y=x$ for $k=100$ where $z=x+iy$. 
This seems to be consistent with the perturbative result obtained in section \ref{perturbative}. 
In fact, the solution at order $\lambda^0$ describes a distribution of $\phi_i$ which are on the line $y=x$. 
On the other hand, for $k=1$, the value of $x$ increases compared with the $k=100$ case, 
but the value of $y$ does not increase as much as $x$, and therefore the solution $z$ is 
placed apart from the line $y=x$. 

Based on this simple observation, it is tempting to speculate that the distribution of $\phi_i$ may become broadly 
extended in the $x$-direction, while in the $y$-direction the width of the distribution would be of order $\lambda^0$ 
in the large $N$ limit with $\lambda$ large but finite. 
In the next section, we will argue that the appropriate rescaling of the variables would be 
\begin{equation}
\pi\phi_i = \Lambda(\lambda) x_i+if(x_i), 
   \label{scaling}
\end{equation}
where $\Lambda(\lambda)$ is an increasing function of $\lambda$, and $f(x)$ is a function of order $\lambda^0$. 
The new variables $x_i$ are assumed to have a distribution function $\rho(x)$ in the large $N$ limit which is another function of 
order $\lambda^0$. 
These two unknown functions $\rho(x)$ and $f(x)$ are expected to be determined by the real part and the imaginary part of 
(\ref{saddle1}). 
The function $\Lambda(\lambda)$ will be determined by the consistency of the order of $\lambda$. 

The assumption (\ref{scaling}) seems to be appropriate by the following reason. 
Notice that the right-hand side of (\ref{saddle1}) has the term 
\begin{equation}
-\tanh[\pi(\phi_i-\phi_i^*)], 
\end{equation}
since the second sum does not exclude $j=i$. 
Due to this term, Im$(\phi_i)$ cannot approach $\pm\frac14$ with a finite value of Re$(\phi_i)$ since, if this is the case, 
the right-hand side diverges 
while the left-hand side is finite. 
In the large $N$ limit, $\phi_i$ are expected to form a continuous curve $I$ in ${\bf C}$. 
The above observation suggests that $I$ should be confined in a region $\{ z\in{\bf C} | -\frac14<\mbox{Im}(z)<+\frac14 \}$.

\vspace{1cm}

\section{Toward large $\lambda$} \label{largelambda}

\vspace{5mm}

In this section, we try to extract information on the distribution of $\phi_i$ for large $\lambda$ as much as possible. 
With the ansatz $\tilde{\phi}_i=\phi_i^*$, the saddle-point equations (\ref{saddle1})(\ref{saddle2}) reduce to 
\begin{equation}
-\frac i\lambda\phi_i = \frac1N\sum_{j\ne i}\coth[\pi(\phi_i-\phi_j)]-\frac1N\sum_{j}\tanh[\pi(\phi_i-\phi_j^*)]. 
   \label{redABJM}
\end{equation}
This equation is still complicated. 
Let us start with the discussion on a simpler equation, and then gradually increase the complexity of equations. 

\vspace{5mm}

\subsection{coth-model}

\vspace{5mm}

The first equation\footnote{
This equation is actually the saddle-point equation for a matrix model describing Chern-Simons theory on $S^3$ \cite{Marino:2002fk}. 
This matrix model was further investigated in \cite{Aganagic:2002wv}. 
See also \cite{Halmagyi:2003ze}. 
} we consider is 
\begin{equation}
\frac c\lambda \phi_i = \frac1N\sum_{j\ne i}\coth(\phi_i-\phi_j), 
   \label{simple1}
\end{equation}
where $c>0$ is a constant of order $\lambda^0$. 
The distribution of $\phi_i$ is symmetric with respect to the origin $\phi=0$ due to the symmetry of the equation. 
Suppose that the width of the distribution is proportional to $\Lambda(\lambda)$ which satisfies 
\begin{equation}
\lim_{\lambda\to\infty}\Lambda(\lambda) = \infty, 
\end{equation}
and the proportionality coefficient is of order $\lambda^0$. 
This assumption seems natural when (\ref{simple1}) is regarded as an equation for the balance between an external 
confining force acting on 
$\phi_i$ (left-hand side) and a repulsive force among $\phi_i$ (right-hand side). 
As $\lambda$ becomes large, the external force which confines $\phi_i$ around $\phi=0$ becomes small while the repulsive force is 
kept intact. 
As a result, the distribution of $\phi_i$ becomes broader as $\lambda$ becomes larger. 

Define rescaled variables 
\begin{equation}
x_i := \Lambda^{-1}\phi_i. 
\end{equation}
By definition, the width of the distribution of $x_i$ is of order $\lambda^0$. 
Consider the large $N$ limit while keeping $\lambda$ large but finite. 
Then (\ref{simple1}) can be written as 
\begin{equation}
\frac{c\Lambda}{\lambda} x = \int\hspace{-3.5mm}-\hspace{1mm}dx'\,\rho(x')\coth[\Lambda(x-x')], 
   \label{coth}
\end{equation}
in terms of the distribution function $\rho(x)$ of $x_i$. 
We denote the support of $\rho(x)$ as $I$. 
As mentioned above, $I$ should be of the form 
\begin{equation}
I = [-x_m,x_m]\subset{\bf R}, \hspace{5mm} x_m = O(\lambda^0). 
\end{equation}
Eq.(\ref{coth}) can be rewritten as follows, 
\begin{equation}
\frac{c\Lambda}{\lambda}x = \int dx'\,\rho(x')\varepsilon(x-x')
 +\int\hspace{-3.5mm}-\hspace{1mm}dx'\,\rho(x')\frac{e^{-\Lambda|x-x'|}}{\sinh[\Lambda(x-x')]}, 
   \label{coth2}
\end{equation}
where $\varepsilon(x)$ is the sign function. 

Notice that the integrand of the second term in the right-hand side of (\ref{coth2}) is localized in a region 
around $x$ with width of order $\Lambda^{-1}$. 
Therefore, only some local information on $\rho(x)$ determines the dominant part of the integral for large $\lambda$. 
Based on this observation, let us assume that there is a neighborhood $U(x)\subset I$ 
of $x$ in which $\rho(x)$ is smooth, and its Taylor expansion at $x$ has coefficients 
of order $\lambda^0$. 
Under this assumption, the integral is estimated as follows. 
First, consider an integral 
\begin{equation}
\int_a^b\hspace{-5mm}-\hspace{2mm}dx\,\frac{x^ne^{-\Lambda|x|}}{\sinh \Lambda x} 
 = \int_{-\infty}^{+\infty}\hspace{-8.75mm}-\hspace{5mm}dx\,\frac{x^ne^{-\Lambda|x|}}{\sinh \Lambda x}
  -\int_{-\infty}^adx\,\frac{x^ne^{-\Lambda|x|}}{\sinh \Lambda x}
  -\int_b^{+\infty}dx\,\frac{x^ne^{-\Lambda x}}{\sinh \Lambda x}, 
    \label{estimate}
\end{equation}
where $a<0<b$ and $n\ge0$. 
The third term in the right-hand side is estimated as 
\begin{equation}
\int_b^{+\infty}dx\,\frac{x^ne^{-\Lambda x}}{\sinh \Lambda x} 
 = b^n\Lambda^{-1}e^{-2\Lambda b}+O(\Lambda^{-2}e^{-2\Lambda b}). 
\end{equation}
The second term is also estimated similarly. 
Both of them are negligible compared with the first term for large $\lambda$. 
Therefore, the integral (\ref{estimate}) is 
\begin{equation}
\int_a^b\hspace{-5mm}-\hspace{2mm}dx\,\frac{x^ne^{-\Lambda|x|}}{\sinh \Lambda x} \sim \left\{
\begin{array}{cc}
\Lambda^{-n-1}n!\,2^{-n}\zeta(n+1) & (n: \mbox{odd}), \\
0, & (n: \mbox{even}). 
\end{array}
\right.
\end{equation}
This estimate allows one to obtain 
\begin{equation}
\int\hspace{-3.5mm}-\hspace{1mm}dx'\,\rho(x')\frac{e^{-\Lambda|x-x'|}}{\sinh[\Lambda(x-x')]} 
 \sim -\frac{\pi^2}{12}\Lambda^{-2}\rho'(x). 
\end{equation}
This term is actually negligible compared with the first term in the right-hand side of (\ref{coth2}). 
For large $\lambda$, $\rho(x)$ satisfies 
\begin{equation}
\frac{c\Lambda}{\lambda}x \sim \int dx'\,\rho(x')\varepsilon(x-x'). 
\end{equation}
This implies 
\begin{equation}
\Lambda(\lambda) = \lambda, \hspace{5mm} \rho(x) \sim \frac c2. 
\end{equation}

The above analysis fails where $x$ is very close to one of 
the boundaries of $I$ so that the width of $U(x)$ is less than or of 
the same order of $\Lambda^{-1}$, and where $\rho(x)$ starts changing rapidly. 
The simplest possible solution for $\rho(x)$ would be such a function which is almost constant except for regions 
$|x-x_m|=O(\Lambda^{-1})$ and $|x+x_m|=O(\Lambda^{-1})$, and which rapidly decreases to zero near the 
boundaries of $I$. 
For this solution, $x_m$ is determined to be 
\begin{equation}
x_m \sim c^{-1}
\end{equation}
for large $\lambda$. 

The results obtained so far turn out to be enough to deduce an approximate solution of (\ref{coth}) which may converge 
to the exact solution in the large $\lambda$ limit. 
The approximate solution is 
\begin{equation}
\rho(x) = \left\{ 
\begin{array}{cc}
-\frac c2 \tanh[\Lambda(x+c^{-1})]\tanh[\Lambda(x-c^{-1})], & (|x|\le c^{-1}) \\ 0, & (|x|>c^{-1}). 
\end{array}
\right. 
   \label{rho1}
\end{equation}
Let $f(x)$ be defined as 
\begin{equation}
f(x) := \int_{-c^{-1}}^{+c^{-1}}\hspace{-10.5mm}-\hspace{5mm}dx'\,\rho(x')\coth[\Lambda(x-x')]. 
\end{equation}
If $x$ is not equal to $\pm c^{-1}$, then the argument above is valid for a large enough $\lambda$, and therefore, 
(\ref{coth}) is satisfied. 
On the other hand, if $x=+c^{-1}$, then $f(c^{-1})$ becomes 
\begin{equation}
f(c^{-1}) = \frac c2\int_{-c^{-1}}^{+c^{-1}}\hspace{-10.5mm}-\hspace{5mm}dx'\,\tanh[\Lambda(x'+c^{-1})] 
\end{equation}
which satisfies (\ref{coth}) in the large $\lambda$ limit. 
The case $x=-c^{-1}$ is similar. 

\begin{figure}
\epsfxsize=2.2in
\begin{center}
\includegraphics[scale=0.5]{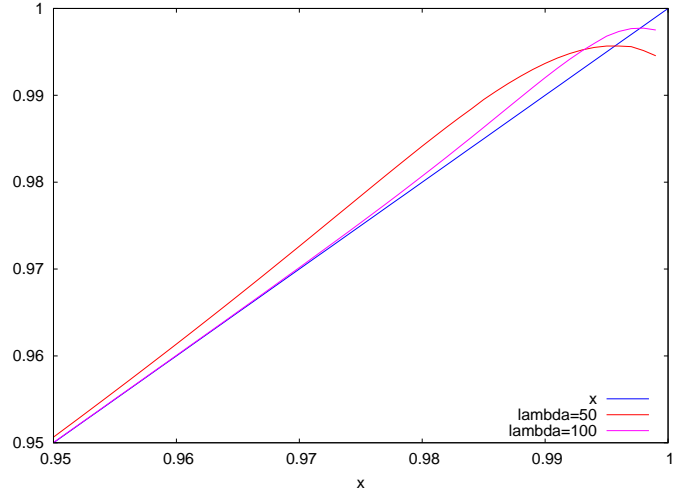}
\end{center}
\caption{Plot of $f(x)$ with $\Lambda=50,100$. $f(x)$ approaches to $x$ as $\Lambda$ becomes large. }
\label{plot coth}
\end{figure}

The graphs of $f(x)$ with $c=1$, 
which is expected to be a good approximation to $x$, are plotted in Fig.\ref{plot coth} for $\Lambda=50,100$. 
The graphs indicate that (\ref{rho1}) would be a good approximation in the large $\lambda$ limit. 

\vspace{5mm}

We have found that the distribution of $\phi_i$ satisfying (\ref{simple1}) has a width $2c^{-1}\lambda$ for large $\lambda$. 
This width is larger than the one in a model where 
$\coth x$ in the right-hand side is replaced with $x^{-1}$, that is, in the case of the 
Gaussian matrix model. 
This is simply because $\coth x$ provides an infinite-range repulsive force. 

In fact, the above information on $\rho(x)$ is enough to determine the large $\lambda$ behavior of the ``Wilson loop'' 
$W_1(\lambda)$: 
\begin{eqnarray}
W_1(\lambda) 
&:=& \int_{-c^{-1}}^{+c^{-1}}dx\,\rho(x)e^{\lambda x} \nonumber \\
&\sim& e^{c^{-1}{\lambda}}. 
\end{eqnarray}

A derivation of this estimate is as follows. 
Consider the estimate of 
\begin{equation}
W(\Lambda) = \int_{-x_m}^{+x_m}dx\,\rho(x)e^{\Lambda x}
\end{equation}
for large $\Lambda$. 
This can be written as 
\begin{equation}
W(\Lambda) = e^{\Lambda x_m}\int_0^{2x_m}du\,\rho(x_m-u)e^{-\Lambda u}. 
\end{equation}
Since $\Lambda$ is large, only the information of $\rho(x_m-u)$ around $u=0$ is important. 
Suppose that $\rho(x_m-u)$ behaves as $u^\alpha\tilde{\rho}(u)$ where $\tilde{\rho}(u)$ is regular at $u=0$. 
Then, the leading term of the asymptotic expansion of $W(\lambda)$ with respect to $\Lambda^{-1}$ is 
\begin{eqnarray}
W(\Lambda) 
&\sim& e^{\Lambda x_m}\int_0^\infty du\,u^\alpha\tilde{\rho}(0) e^{-\Lambda u} \nonumber \\
&=& \tilde{\rho}(0)\Gamma(\alpha+1)\Lambda^{-\alpha-1}e^{\Lambda x_m}. 
\end{eqnarray}

It should be noted that the coth-model was solved exactly in the large $N$ limit in \cite{Aganagic:2002wv}\cite{Halmagyi:2003ze}. 
The resulting distribution function has the behavior which we found in the above analysis. 

\vspace{5mm}

\subsection{cosech-model}

\vspace{5mm}

Next, let us consider 
\begin{eqnarray}
\frac c\lambda \phi_i 
&=& \frac1N\sum_{j\ne i}\coth(\phi_i-\phi_j)-\frac1N\sum_j\tanh(\phi_i-\phi_j) \nonumber \\
&=& \frac1N\sum_{j\ne i}\frac2{\sinh[2(\phi_i-\phi_j)]}. 
   \label{cosech}
\end{eqnarray}
In this equation, the infinite-range repulsive force due to $\coth x$ terms is canceled by $\tanh x$ terms. 
This equation looks closer to (\ref{redABJM}). 

As in the previous subsection, define rescaled variables 
\begin{equation}
x_i := \Lambda^{-1}\phi_i
\end{equation}
so that $x_i$ are always of order $\lambda^0$. 
In the large $N$ limit, (\ref{cosech}) becomes  
\begin{equation}
\frac{c\Lambda}{\lambda}x = \int\hspace{-3.5mm}-\hspace{1mm}dx'\,\frac{2\rho(x')}{\sinh[2\Lambda(x-x')]}. 
   \label{simple2}
\end{equation}
As in the coth-model, there is an integral in the right-hand side 
whose integrand is localized around $x$ for large $\lambda$. 
Therefore, the integral can be estimated similarly. 
One obtains 
\begin{equation}
\int\hspace{-3.5mm}-\hspace{1mm}dx'\,\frac{2\rho(x')}{\sinh[2\Lambda(x-x')]} 
 \sim -\frac{\pi^2}4\Lambda^{-2}\rho'(x). 
   \label{estimate cosech}
\end{equation}
A difference from the coth-model is that there is no $O(\lambda^0)$ term in the right-hand side of (\ref{simple2}). 
As a result, the right-hand side must be small for large $\lambda$, implying a different functional form of $\Lambda(\lambda)$. 
The appropriate choice of $\Lambda$ in this case is 
\begin{equation}
\Lambda(\lambda) = \lambda^{\frac13}. 
\end{equation}
This is a reasonale result. 
The width of eigenvalues is smaller than that of both the coth-model and the Gaussian model, reflecting the fact that the range of 
the repulsive force among the eigenvalues in the cosech-model is the shortest. 

The equations (\ref{simple2}) and (\ref{estimate cosech}) determine $\rho(x)$ as 
\begin{equation}
\rho(x) \sim \frac{2c}{\pi^2}(x_m^2-x^2)
    \label{approx}
\end{equation}
where $x_m$ is an integration constant. 
It turns out that $x_m$ is determined by requiring 
\begin{equation}
\int_{-x_m}^{+x_m}dx\,\frac{2c}{\pi^2}(x_m^2-x^2) = 1. 
    \label{cosec x_m}
\end{equation}

The analysis of the coth-model suggests that an approximate 
solution for the large $\lambda$ limit would be obtained by modifying $\rho(x)$ 
slightly at the boundary of its support such that the modified distribution function satisfies (\ref{simple2}) at $x=\pm x_m$. 
A possible modification is 
\begin{equation}
\rho(x) = -\frac{2c}{\pi^2}\left[ x_m^2-x^2+\frac\xi\Lambda \right]\tanh[2\Lambda(x-x_m)]\tanh[2\Lambda(x+x_m)], 
   \label{rho cosech}
\end{equation}
where 
\begin{equation}
\xi = \pi-\frac4\pi\sum_{n=0}^\infty\frac{(-1)^n}{(2n+1)^2}. 
\end{equation}
This function $\rho(x)$ satisfies the correct normalization condition due to (\ref{cosec x_m}) in the large $\lambda$ limit. 
Note that the addition of the constant $\frac\xi\Lambda$, since this is small, does not contradict with the result (\ref{approx}). 
The constant $\xi$ is chosen such that $\lim_{\lambda\to\infty}g(x_m)=cx_m$ is satisfied, where $g(x)$ is defined as 
\begin{equation}
g(x) := \Lambda^2\int_{-x_m}^{+x_m}\hspace{-10mm}-\hspace{5mm}dx'\,\frac{2\rho(x)}{\sinh[2\Lambda(x-x')]}. 
\end{equation}

\begin{figure}
\epsfxsize=2.2in
\begin{center}
\includegraphics[scale=0.5]{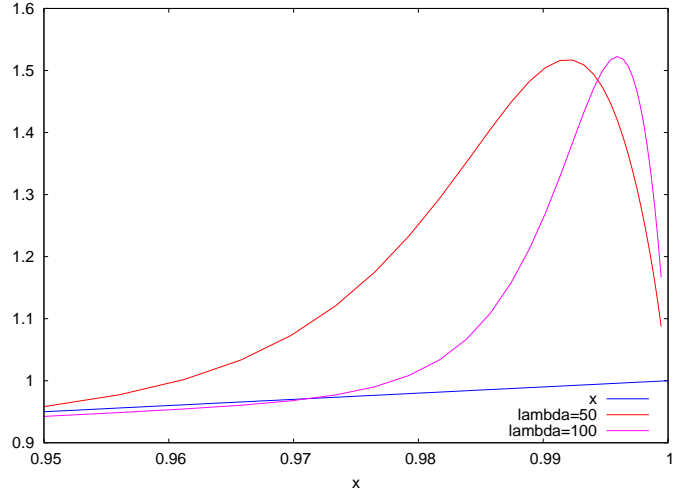}
\end{center}
\caption{Plot of $g(x)$ with $\Lambda=50,100$. $g(x)$ approaches $x$ point-wise.}
\label{plot cosech}
\end{figure}

The graphs of $g(x)$ with $x_m=1$ and $c=1$, are plotted in Fig.\ref{plot cosech} for $\Lambda=50,100$. 
By the argument using the local nature of the integrand, one can show that $g(x)$ converges to $x$ point-wise in the large 
$\lambda$ limit. 

The large $\lambda$ behavior of the Wilson loop $W_2(\lambda)$ in this case is 
\begin{eqnarray}
W_2(\lambda) 
&:=& \int_{-x_m}^{+x_m}dx\,\rho(x)e^{\lambda^\frac13 x} \nonumber \\
&\sim& e^{x_m{\lambda}^\frac13}. 
\end{eqnarray}

\vspace{5mm}

The Fig.\ref{plot cosech}, however, 
suggests that the maximum deviation of $g(x)$ from $x$, or sup$_{x\in [-1,1]}|g(x)-x|$, might not vanish in the large $\lambda$ 
limit. 
Indeed, it can be shown that 
\begin{equation}
\lim_{\Lambda\to\infty} g(x) 
 = \frac{2}{\pi^2}\int_{-\infty}^y\hspace{-7.25mm}-\hspace{5mm}du\,\frac{\tanh(u-y)}{\sinh u}(y-u+\xi)
\end{equation}
holds, where $y=2\Lambda(x_m-x)$ is assumed to be of order $\Lambda^0$. 
This implies that 
\begin{equation}
\lim_{\Lambda\to\infty}\sup_{x\in [-1,1]}|g(x)-x|>0. 
\end{equation}
As a result, the convergence of $g(x)$ to $x$ is not uniform. 
This may not be a serious drawback if one is only interested in the large $\lambda$ limit. 
If one would like to discuss also a $1/\lambda$ corrections, it might be better to start with another approximate solution. 

\vspace{5mm}

\subsection{ABJM equation}

\vspace{5mm}

The experiences in analyzing the previous models will be helpful to consider the saddle-point equation for ABJM theory. 
Recall that the saddle-point equation we would like to study is 
\begin{equation}
-\frac i{\lambda}\phi_i = \frac1N\sum_{j\ne i}\coth[\pi(\phi_i-\phi_j)]-\frac1N\sum_{j}\tanh[\pi(\phi_i-\phi_j^*)]. 
    \label{ABJM}
\end{equation}

In the following, we assume the existence of a solution compatible with the following scaling behavior 
\begin{equation}
\phi_i = \frac1\pi[\Lambda x_i+if(x_i)], 
    \label{ABJMscaling}
\end{equation}
where $x_i\in{\bf R}$ and $f(x)$ is a real function of order $\lambda^0$. 
In the large $N$ limit, (\ref{ABJM}) becomes   
\begin{eqnarray}
-\frac i\pi\left[ \frac{\Lambda}{\lambda}x+\frac i\lambda f(x) \right] 
&=& \int \hspace{-3.5mm}-\hspace{1mm}dx'\, \rho(x') \coth[\Lambda(x-x')+i(f(x)-f(x'))] \nonumber \\
& & -\int dx'\,\rho(x')\tanh[\Lambda(x-x')+i(f(x)+f(x'))]. 
  \label{ABJM2}
\end{eqnarray}
This consists of two independent equations which determine two real functions $\rho(x)$ and $f(x)$. 

As in the previous subsections, the integrand in the right-hand side is localized around $x$ for large $\lambda$. 
Explicitly, 
\begin{eqnarray}
\coth(a+ib_-) &=& \tanh a+\frac1{\cosh^2a}\frac1{\tanh a+i\tan b_-}, \\
\tanh(a+ib_+) &=& \tanh a+\frac1{\cosh^2a}\frac{i\tan b_+}{1+i\tanh a\tan b_+}, 
\end{eqnarray}
where $a=\Lambda(x-x')$ and $b_\pm=f(x)\pm f(x')$, and the infinite-range part $\tanh a$ cancels, as in the cosech-model. 
Therefore, assuming the smoothness of $\rho(x)$ and $f(x)$ in a neighborhood $U(x)\subset {\bf R}$ of $x$, 
only local information on those 
functions determines the integrals. 

First, we separate the complex equation (\ref{ABJM2}) into its real part and imaginary part. 
It is possible to 
show that the expansion in terms of $b_-$ provides an asymptotic expansion of the corresponding integral with 
respect to $\Lambda^{-1}$. 
In addition, the expansion in terms of $(\tan b_-/\tanh a)$ also provides sub-leading terms in $\Lambda^{-1}$, due to the locality 
of the integrand. 
Taking into account only the leading contributions to the asymptotic expansion, (\ref{ABJM2}) is approximated as 
\begin{eqnarray}
\frac1{\pi\lambda}f(x) 
&\sim& \int \hspace{-3.5mm}-\hspace{1mm}dx'\, \left[ \frac{2\rho(x')}{\sinh[2\Lambda(x-x')]}
   +\frac1{\coth^2[\Lambda(x-x')]}\frac{\tanh[\Lambda(x-x')]\tan^2(2f(x))}
    {1+\tanh^2[\Lambda(x-x')]\tan^2(2f(x))} \right] \nonumber 
   \label{ABJM3-1} \\
&\sim& -\frac{\pi^2}4\Lambda^{-2}\rho'(x), \\
\frac{\Lambda}{\pi\lambda}x 
&\sim& \int \hspace{-3.5mm}-\hspace{1mm}dx'\, \rho(x')\left[ \frac{f(x)-f(x')}{\sinh^2[\Lambda
    (x-x')]}+\frac1{\coth^2[\Lambda(x-x')]}\frac{\tan(2f(x))}{1+\tanh^2[\Lambda(x-x')]\tan^2(2f(x))}  \right] \nonumber \\
&\sim& 4\Lambda^{-1}\rho(x)f(x). 
   \label{ABJM3-2}
\end{eqnarray}
Here we assumed $|f(x)|<\frac \pi4$. 
Note that naively the order of $\Lambda$ of the second integral in (\ref{ABJM3-1}) is $\Lambda^{-1}$, but the contribution of 
this order vanishes. 
These equations are consistent with the scaling assumption (\ref{ABJMscaling}) iff 
\begin{equation}
\Lambda(\lambda)=\sqrt{\lambda}. 
\end{equation}
$\rho(x)$ and $f(x)$ are determined by the above equations. 
The result is 
\begin{eqnarray}
\rho(x) &\sim& \frac1{\pi^2}\sqrt{2c^2-x^2}, 
   \label{soln1} \\
f(x) &\sim& \frac\pi4\frac x{\sqrt{2c^2-x^2}}, 
   \label{soln2}
\end{eqnarray}
where $c>0$ is an integration constant. 

Note that the saddle-point equations are consistent with our ansatz for $f(x)$ to be confined within $[-\frac\pi 4,+\frac\pi4]$. 
In the case $\frac\pi4<f(x)<\frac\pi2$, for example, (\ref{ABJM3-2}) is replaced with 
\begin{equation}
\frac\Lambda{\pi\lambda}x \sim \Lambda^{-1}\rho(x)(4f-2\pi). 
\end{equation}
This implies $\rho(x)<0$, strongly suggesting that the solutions (\ref{soln1})(\ref{soln2}) do not extend to a region where 
$|f(x)|>\frac\pi4$, 
that is, $\rho(x)$ is non-zero only if $|x|\le c$. 
Since $\rho(c)>0$, this implies that $\rho(x)$ suddenly starts decreasing to zero at a point in the interval $[0,c]$, 
just like the coth-model 
discussed before. 

Let $x_m\le c$ be the position of the right-edge of the support of $\rho(x)$. 
$x_m$ is determined by requiring 
\begin{equation}
\int_{-x_m}^{+x_m}dx\,\rho(x) = 1, 
   \label{density normalization}
\end{equation}
provided that $c$ is given by another input. 
Since we do not know at present how to determine $c$, the exact value of $x_m$ cannot be obtained from the analysis 
explained so far. 
The best we can do is to put an upper bound on $x_m$ which is derived as follows. 
The approximate form (\ref{soln1}) of $\rho(x)$ indicates that $x_m$ becomes large if $c$ becomes small. 
Therefore, under the condition $x_m\le c$, the largest value of $x_m$ determined from 
(\ref{density normalization}) is obtained when $c$ is chosen to be equal to $x_m$. 
For this particular choice, (\ref{density normalization}) determines $c$ to be 
\begin{equation}
c = \sqrt{\frac{2\pi^2}{\pi+2}}. 
   \label{coefficient}
\end{equation}
This turns out to be the minimum possible value of the integration constant $c$ suitable for our problem\footnote{
In some earlier versions of the paper, we claimed that the value (\ref{coefficient}) of $c$ would be equal to $x_m$. 
After the appearance of \cite{Marino:2009jd}, the arguments were re-examined and then 
it turned out that this claim was too strong which could not be justified by the analysis 
done in this section. 
As explained above, the best we can find is the upper bound on $x_m$ which is compatible with 
\cite{Marino:2009jd}. 
}. 
This gives the upper bound 
\begin{equation}
x_m \le \sqrt{\frac{2\pi^2}{\pi+2}}. 
   \label{upper bound}
\end{equation}
The behavior of $W(\lambda)$ in the large $\lambda$ limit is then expected to be 
\begin{equation}
|W(\lambda)| \sim e^{2\pi x_m\sqrt{\lambda}}. 
\end{equation}
This exponential behavior and the upper bound (\ref{upper bound}) on the coefficient are consistent with the conjectured 
correspondence between the Wilson loop and a string worldsheet in AdS$_4\times {\bf CP}^3$ 
proposed in \cite{Drukker:2008zx}\cite{Chen:2008bp}\cite{Rey:2008bh}. 
Especially, the appearance of $\sqrt{\lambda}$ in the exponent strongly suggests the existence of a dual string worldsheet 
in $AdS_4$.

\vspace{1cm}

\section{Discussion} \label{discuss}

\vspace{5mm}

We have investigated the saddle-point equations (\ref{saddle1})(\ref{saddle2}). 
For small $\lambda$, we found that the equations reproduce the perturbative expansion of the 
expectation value $W(\lambda)$ of the Wilson loop (\ref{BPSWL}). 
We also found an efficient algorithm to determine the perturbative expansion to any desired order of $\lambda$. 
For large $\lambda$, we found an approximate solution to the saddle-point equations which is expected to converge to the 
exact solution in the large $\lambda$ limit. 
Based on the solution, we determined the large $\lambda$ behavior of $W(\lambda)$ which may be compatible with AdS/CFT 
correspondence. 

The algorithm found in section \ref{algorithm} is rather simple. 
This could provide a recursive formula for $\langle \hat{x}^k \rangle$, and therefore for $W(\lambda)$. 
It is very interesting to analytically solve this recursive formula to obtain a closed formula for $W(\lambda)$. 
It is already interesting if one can 
determine all the coefficients of the perturbative expansion using the recursive formula since 
some non-perturbative information may be extracted from them. 

The understanding of the solution to (\ref{saddle1})(\ref{saddle2}) for large $\lambda$ obtained in this paper is still at the 
primitive level. 
Especially, the explicit form of $\rho(x)$ may have ambiguities at finite $\lambda$ which cannot be fixed by the argument of 
locality and the large $\lambda$ limit. 
For example, in the cosech-model, although the expression (\ref{rho cosech}) 
provides an approximate solution which converges to the exact solution 
in the large $\lambda$ limit, it is not clear whether it is 
appropriate to use the expression to discuss $1/\lambda$ corrections. 
It is quite important to obtain the solution for finite $\lambda$ which can interpolate the small $\lambda$ result and the 
large $\lambda$ result. 
One may expect that the holomorphicity of 
\begin{equation}
{\cal R}(z) := \int_Idx\,\rho(x)\coth[\pi(x-z)]
\end{equation}
as well as the conditions derived from the saddle-point equations might be enough to determine $\rho(x)$ uniquely. 
Indeed, ${\cal R}(z)$ turns out to satisfy some equations which are similar to the one discussed in 
\cite{Kazakov:1998ji}. 
In the case of ABJM theory, the periodicity of ${\cal R}(z)$ in the imaginary direction makes the problem complicated. 
It is very interesting to solve Eqs.(\ref{saddle1})(\ref{saddle2}) exactly, for example, in this way. 
Another way to gain some information on the large $\lambda$ solution is to solve (\ref{saddle1})(\ref{saddle2}) 
numerically for a large $N$, 
extending the analysis of section \ref{finiteN}. 

Since the localization of \cite{Kapustin:2009kz} 
can be applied to a quite general family of supersymmetric Chern-Simons-matter theories, 
one can extend the analysis exhibited in this paper to more general theories. 
A systematic research in this direction may shade some light on the understanding of AdS$_4$/CFT$_3$ correspondence.

\vspace{2cm}

{\bf \Large Acknowledgements}

\vspace{5mm}

We would like to thank Seok Kim, Dongmin Gang and Eunkyung Koh for valuable discussions and comments. 
This work was supported by the BK21 program of the Ministry of Education, Science and Technology, 
National Science Foundation of Korea Grants R01-2008-000-10656-0, 
2005-084-C00003, 2009-008-0372 and EU-FP Marie Curie Research 
\& Training Network HPRN-CT-2006-035863 (2009-06318).

\appendix

\vspace{2cm}

{\bf \Large Appendix}

\vspace{1cm}

\section{Calculation at order $\lambda^{10}$} \label{higher_order}

\vspace{5mm}

The result is the following: 
The parameters in the resolvent $R_{10}(x)$ are 
\begin{eqnarray}
\alpha_1 
&=& {{95034445994244677\,i\,\pi^9\,\lambda^{10}}\over{11693788644037500}}+{{824385881240071 \,\pi^8\,\lambda^9}\over{155917181920500}}-{{1792539152\,i\,\pi^7\,\lambda^8 }\over{516891375}}-{{8481442\,\pi^6\,\lambda^7}\over{3648645}} \nonumber \\
& & +{{48010852 \,i\,\pi^5\,\lambda^6}\over{30405375}}+{{34774\,\pi^4\,\lambda^5}\over{31185}}- {{2288\,i\,\pi^3\,\lambda^4}\over{2835}}-{{604\,\pi^2\,\lambda^3}\over{945}}+{{8 \,i\,\pi\,\lambda^2}\over{15}}+{{2\,\lambda}\over{3}}-{{i}\over{\pi}}, \\
\alpha_2 &=& -{{1300145665611332\,\pi^8\,\lambda^{10}}\over{27842353914375}}+{{344633417596672 \,i\,\pi^7\,\lambda^9}\over{12993098493375}}+{{186341275429\,\pi^6\,\lambda^8 }\over{12405393000}}-{{7321664\,i\,\pi^5\,\lambda^7}\over{868725}} \nonumber \\
& &-{{4084264 \,\pi^4\,\lambda^6}\over{868725}}+{{239872\,i\,\pi^3\,\lambda^5}\over{93555}}+{{3907 \,\pi^2\,\lambda^4}\over{2835}}-{{128\,i\,\pi\,\lambda^3}\over{189}}-{{14\,\lambda^2 }\over{45}}, \\
\alpha_3 &=& -{{441275989540096\,i\,\pi^7\,\lambda^{10}}\over{4218538471875}}-{{75094926421943 \,\pi^6\,\lambda^9}\over{1484925542100}}+{{183862829632\,i\,\pi^5\,\lambda^8 }\over{7753370625}}+{{64951384\,\pi^4\,\lambda^7}\over{6081075}} \nonumber \\
& & -{{27517696 \,i\,\pi^3\,\lambda^6}\over{6081075}}-{{482\,\pi^2\,\lambda^5}\over{275}}+{{128 \,i\,\pi\,\lambda^4}\over{225}}+{{124\,\lambda^3}\over{945}}, \\
\alpha_4 &=& {{2535162760294933\,\pi^6\,\lambda^{10}}\over{21417195318750}}-{{87217425750016 \,i\,\pi^5\,\lambda^9}\over{1856156927625}}-{{151498931417\,\pi^4\,\lambda^8 }\over{8683775100}}+{{752263168\,i\,\pi^3\,\lambda^7}\over{127702575}} \nonumber \\
& & +{{73397228 \,\pi^2\,\lambda^6}\over{42567525}}-{{4096\,i\,\pi\,\lambda^5}\over{10395}}-{{254 \,\lambda^4}\over{4725}}, \\
\alpha_5 &=& {{30838712389203968\,i\,\pi^5\,\lambda^{10}}\over{417635308715625}}+{{4653424941601 \,\pi^4\,\lambda^9}\over{202489846650}}-{{11226873856\,i\,\pi^3\,\lambda^8 }\over{1776226725}}-{{50380768\,\pi^2\,\lambda^7}\over{34827975}} \nonumber \\
& & +{{2830336 \,i\,\pi\,\lambda^6}\over{11609325}}+{{292\,\lambda^5}\over{13365}}, \\
\alpha_6 &=& -{{6113412629359571\,\pi^4\,\lambda^{10}}\over{235589148506250}}+{{230520782848 \,i\,\pi^3\,\lambda^9}\over{38940355125}}+{{43219333\,\pi^2\,\lambda^8}\over{39760875 }}-{{65536\,i\,\pi\,\lambda^7}\over{467775}}-{{5657908\,\lambda^6}\over{638512875 }}, \\
\alpha_7 &=& -{{353963361107968\,i\,\pi^3\,\lambda^{10}}\over{70676744551875}}-{{4193522922926 \,\pi^2\,\lambda^9}\over{5568470782875}}+{{118521856\,i\,\pi\,\lambda^8}\over{1550674125 }}+{{65528\,\lambda^7}\over{18243225}}, \\
\alpha_8 &=& {{927244437672232\,\pi^2\,\lambda^{10}}\over{1891642280653125}}-{{91995766784 \,i\,\pi\,\lambda^9}\over{2292899734125}}-{{33862354\,\lambda^8}\over{23260111875 }}, \\
\alpha_9 &=& {{183092903936\,i\,\pi\,\lambda^{10}}\over{8955143071875}}+{{22998766228 \,\lambda^9}\over{38979295480125}}, \\
\alpha_{10} &=& -{{366185109428\,\lambda^{10}}\over{1531329465290625}}, \\
\beta_1 &=& {{25177793483209721\,i\,\pi^9\,\lambda^{10}}\over{20581068013506000}}+{{9640092302437393 \,\pi^8\,\lambda^9}\over{6236687276820000}}-{{3455382824\,i\,\pi^7\,\lambda^8 }\over{5746615875}}-{{3032042257\,\pi^6\,\lambda^7}\over{3831077250}} \nonumber \\
& & +{{3237679 \,i\,\pi^5\,\lambda^6}\over{9823275}}+{{436883\,\pi^4\,\lambda^5}\over{935550}}- {{622\,i\,\pi^3\,\lambda^4}\over{2835}}-{{346\,\pi^2\,\lambda^3}\over{945}}+{{2 \,i\,\pi\,\lambda^2}\over{9}}+{{2\,\lambda}\over{3}}-{{i}\over{\pi}}, \\
\beta_2 &=& -{{1308690820561018067\,\pi^8\,\lambda^{10}}\over{68603560045020000}}+{{744839342044528 \,i\,\pi^7\,\lambda^9}\over{64965492466875}}+{{6874823870267\,\pi^6\,\lambda^8 }\over{976924698750}}-{{128817718\,i\,\pi^5\,\lambda^7}\over{30405375}} \nonumber \\
& & -{{372482207 \,\pi^4\,\lambda^6}\over{141891750}}+{{734336\,i\,\pi^3\,\lambda^5}\over{467775 }}+{{4618\,\pi^2\,\lambda^4}\over{4725}}-{{172\,i\,\pi\,\lambda^3}\over{315}}-{{14 \,\lambda^2}\over{45}}, \\
\beta_3 &=& -{{17639281196348848\,i\,\pi^7\,\lambda^{10}}\over{306265893058125}}-{{88912240611058 \,\pi^6\,\lambda^9}\over{2998407344625}}+{{186517708697\,i\,\pi^5\,\lambda^8 }\over{12524675625}}+{{1542519221\,\pi^4\,\lambda^7}\over{212837625}} \nonumber \\
& & -{{2134621256 \,i\,\pi^3\,\lambda^6}\over{638512875}}-{{2248\,\pi^2\,\lambda^5}\over{1575}}+{{2434 \,i\,\pi\,\lambda^4}\over{4725}}+{{124\,\lambda^3}\over{945}}, \\
\beta_4 &=& {{5091572820200602367\,\pi^6\,\lambda^{10}}\over{64315837542206250}}-{{2166511481052202 \,i\,\pi^5\,\lambda^9}\over{64965492466875}}-{{12902809900867\,\pi^4\,\lambda^8 }\over{976924698750}}+{{3053312744\,i\,\pi^3\,\lambda^7}\over{638512875}} \nonumber \\
& & + {{963077002\,\pi^2\,\lambda^6}\over{638512875}}-{{6964\,i\,\pi\,\lambda^5}\over{18711 }}-{{254\,\lambda^4}\over{4725}}, \\
\beta_5 &=& {{607486105446138962\,i\,\pi^5\,\lambda^{10}}\over{10719306257034375}}+{{1210297833352637 \,\pi^4\,\lambda^9}\over{64965492466875}}-{{2651143334458\,i\,\pi^3\,\lambda^8 }\over{488462349375}}-{{280467644\,\pi^2\,\lambda^7}\over{212837625}} \nonumber \\
& & +{{50003524 \,i\,\pi\,\lambda^6}\over{212837625}}+{{292\,\lambda^5}\over{13365}}, \\
\beta_6 &=& -{{26234503780513031\,\pi^4\,\lambda^{10}}\over{1191034028559375}}+{{60442600532504 \,i\,\pi^3\,\lambda^9}\over{11464498670625}}+{{23611974296\,\pi^2\,\lambda^8 }\over{23260111875}}-{{355768\,i\,\pi\,\lambda^7}\over{2606175}} \nonumber \\
& & -{{5657908 \,\lambda^6}\over{638512875}}, \\
\beta_7 &=& -{{147061705524924976\,i\,\pi^3\,\lambda^{10}}\over{32157918771103125}}- {{139315307981924\,\pi^2\,\lambda^9}\over{194896477400625}}+{{1743965486\, i\,\pi\,\lambda^8}\over{23260111875}}+{{65528\,\lambda^7}\over{18243225}}, \\
\beta_8 &=& {{201731372996882\,\pi^2\,\lambda^{10}}\over{428772250281375}}-{{1580440276 \,i\,\pi\,\lambda^9}\over{39978764595}}-{{33862354\,\lambda^8}\over{23260111875 }}, \\
\beta_9 &=& {{30942701463628\,i\,\pi\,\lambda^{10}}\over{1531329465290625}}+{{22998766228 \,\lambda^9}\over{38979295480125}}, \\
\beta_{10} &=& -{{366185109428\,\lambda^{10}}\over{1531329465290625}}, \\
\gamma &=& -{{161894160651233\,i\,\pi^{11}\,\lambda^{10}}\over{24198786612000}}+{{535693251913 \,\pi^{10}\,\lambda^9}\over{109994484600}}+{{10384955081\,i\,\pi^9\,\lambda^8 }\over{3308104800}}-{{60232\,\pi^8\,\lambda^7}\over{25025}} \nonumber \\
& & -{{5053207\,i\, \pi^7\,\lambda^6}\over{3153150}}+{{13922\,\pi^6\,\lambda^5}\over{10395}}+{{2977 \,i\,\pi^5\,\lambda^4}\over{3150}}-{{296\,\pi^4\,\lambda^3}\over{315}}-{{34\,i\, \pi^3\,\lambda^2}\over{45}}+{{4\,\pi^2\,\lambda}\over{3}}+2\,i\,\pi. 
\end{eqnarray}
$R_{10}(x)$ can be expanded as 
\begin{equation}
R_{10}(x) = \frac1x+\sum_{k=1}^\infty \langle \hat{x}^{2k} \rangle_{10}x^{-2k-1}, 
\end{equation}
where 
\begin{eqnarray}
\langle \hat{x}^{2} \rangle_{10} &=& -{{22485647\,i\,\pi^{11}\,\lambda^{10}}\over{49397040}}+{{227813\,\pi^{10 }\,\lambda^9}\over{453600}}+{{3307813\,i\,\pi^9\,\lambda^8}\over{13608000}}-{{863 \,\pi^8\,\lambda^7}\over{3024}}-{{961\,i\,\pi^7\,\lambda^6}\over{6615}}+{{17\, \pi^6\,\lambda^5}\over{90}} \nonumber \\
& & +{{283\,i\,\pi^5\,\lambda^4}\over{2700}}-{{\pi^4\,\lambda^3 }\over{6}}-{{i\,\pi^3\,\lambda^2}\over{9}}+{{\pi^2\,\lambda}\over{3}}+{{i\,\pi }\over{2}}, \\
\langle \hat{x}^{4} \rangle_{10} &=& {{12000329219\,\pi^{12}\,\lambda^{10}}\over{4322241000}}+{{276624839\,i\, \pi^{11}\,\lambda^9}\over{130977000}}-{{228148609\,\pi^{10}\,\lambda^8}\over{158760000 }}-{{68188\,i\,\pi^9\,\lambda^7}\over{59535}}+{{7799521\,\pi^8\,\lambda^6}\over{9525600 }} \nonumber \\
& & +{{158\,i\,\pi^7\,\lambda^5}\over{225}}-{{1469\,\pi^6\,\lambda^4}\over{2700}}- {{8\,i\,\pi^5\,\lambda^3}\over{15}}+{{29\,\pi^4\,\lambda^2}\over{60}}+{{2\,i\, \pi^3\,\lambda}\over{3}}-{{\pi^2}\over{2}}, \\
\langle \hat{x}^{6} \rangle_{10} &=& {{5323721653441\,i\,\pi^{13}\,\lambda^{10}}\over{449513064000}}-{{413232004 \,\pi^{12}\,\lambda^9}\over{49116375}}-{{37102261243\,i\,\pi^{11}\,\lambda^8 }\over{6286896000}}+{{3448213\,\pi^{10}\,\lambda^7}\over{793800}} \nonumber \\
& & +{{2267093 \,i\,\pi^9\,\lambda^6}\over{714420}}-{{74381\,\pi^8\,\lambda^5}\over{30240}}-{{4811 \,i\,\pi^7\,\lambda^4}\over{2520}}+{{101\,\pi^6\,\lambda^3}\over{63}}+{{41\,i\, \pi^5\,\lambda^2}\over{30}}-{{5\,\pi^4\,\lambda}\over{4}}-{{5\,i\,\pi^3}\over{8 }}, \\
\langle \hat{x}^{8} \rangle_{10} &=& -{{5545927036934549\,\pi^{14}\,\lambda^{10}}\over{121368527280000}}-{{275906240221 \,i\,\pi^{13}\,\lambda^9}\over{8756748000}}+{{44102510327\,\pi^{12}\,\lambda^8 }\over{2020788000}}+{{13002184\,i\,\pi^{11}\,\lambda^7}\over{841995}} \nonumber \\
& & -{{224690939 \,\pi^{10}\,\lambda^6}\over{20412000}}-{{19561\,i\,\pi^9\,\lambda^5}\over{2430}}+{{51349\,\pi^8\,\lambda^4}\over{8640}}+{{1418\,i\,\pi^7\,\lambda^3}\over{315}}- {{203\,\pi^6\,\lambda^2}\over{60}}-{{7\,i\,\pi^5\,\lambda}\over{3}}+{{7\,\pi^4 }\over{8}}, \\
\langle \hat{x}^{10} \rangle_{10} &=& -{{3376146566293313\,i\,\pi^{15}\,\lambda^{10}}\over{20228087880000}}+{{2338727068919 \,\pi^{14}\,\lambda^9}\over{20803910400}}+{{142265159471\,i\,\pi^{13}\,\lambda^8 }\over{1868106240}}-{{4681993579\,\pi^{12}\,\lambda^7}\over{89812800}} \nonumber \\
& & -{{33573821 \,i\,\pi^{11}\,\lambda^6}\over{935550}}+{{16586929\,\pi^{10}\,\lambda^5}\over{665280 }}+{{523711\,i\,\pi^9\,\lambda^4}\over{30240}}-{{24103\,\pi^8\,\lambda^3}\over{2016 }}-{{63\,i\,\pi^7\,\lambda^2}\over{8}}+{{35\,\pi^6\,\lambda}\over{8}}\nonumber \\
& & +{{21\,i\, \pi^5}\over{16}}, \\
\langle \hat{x}^{12} \rangle_{10} &=& {{24761546379007051\,\pi^{16}\,\lambda^{10}}\over{42247941120000}}+{{3010228043441 \,i\,\pi^{15}\,\lambda^9}\over{7801466400}}-{{1081813416673\,\pi^{14}\,\lambda^8 }\over{4245696000}}-{{186678329\,i\,\pi^{13}\,\lambda^7}\over{1105650}} \nonumber \\
& & +{{7915650007 \,\pi^{12}\,\lambda^6}\over{70761600}}+{{307303\,i\,\pi^{11}\,\lambda^5}\over{4158 }}-{{9712351\,\pi^{10}\,\lambda^4}\over{201600}}-{{5753\,i\,\pi^9\,\lambda^3 }\over{189}}+{{1133\,\pi^8\,\lambda^2}\over{64}}+{{33\,i\,\pi^7\,\lambda}\over{4 }} \nonumber \\
& & -{{33\,\pi^6}\over{16}}, \\
\langle \hat{x}^{14} \rangle_{10} &=& {{9588084588732887\,i\,\pi^{17}\,\lambda^{10}}\over{4809475440000}}-{{1547644701254461 \,\pi^{16}\,\lambda^9}\over{1207084032000}}-{{5920625808253\,i\,\pi^{15}\, \lambda^8}\over{7185024000}}+{{23723634703\,\pi^{14}\,\lambda^7}\over{44906400}} \nonumber \\
& &  +{{818216699\,i\,\pi^{13}\,\lambda^6}\over{2432430}}-{{1807931957\,\pi^{12 }\,\lambda^5}\over{8553600}}-{{16755713\,i\,\pi^{11}\,\lambda^4}\over{129600}}+ {{324181\,\pi^{10}\,\lambda^3}\over{4320}}+{{7007\,i\,\pi^9\,\lambda^2}\over{180 }} \nonumber \\
& & -{{1001\,\pi^8\,\lambda}\over{64}}-{{429\,i\,\pi^7}\over{128}}, \\
\langle \hat{x}^{16} \rangle_{10} &=& -{{7920651425068514707\,\pi^{18}\,\lambda^{10}}\over{1200445069824000}}- {{20720319106004807\,i\,\pi^{17}\,\lambda^9}\over{5001854457600}}+{{1461510279337891 \,\pi^{16}\,\lambda^8}\over{564583219200}} \nonumber \\
& & +{{223063873\,i\,\pi^{15}\,\lambda^7 }\over{138996}}-{{1336809970387\,\pi^{14}\,\lambda^6}\over{1362160800}}-{{41814259 \,i\,\pi^{13}\,\lambda^5}\over{71280}}+{{34944793\,\pi^{12}\,\lambda^4}\over{103680 }}+{{19487\,i\,\pi^{11}\,\lambda^3}\over{108}} \nonumber \\
& & -{{40469\,\pi^{10}\,\lambda^2 }\over{480}}-{{715\,i\,\pi^9\,\lambda}\over{24}}+{{715\,\pi^8}\over{128}}, \\
\langle \hat{x}^{18} \rangle_{10} &=& -{{87296746879362569\,i\,\pi^{19}\,\lambda^{10}}\over{4095463680000}}+{{667329042674322019 \,\pi^{18}\,\lambda^9}\over{51111386726400}}+{{5221907328724763\,i\,\pi^{17 }\,\lambda^8}\over{658680422400}} \nonumber \\
& & -{{92034765828557\,\pi^{16}\,\lambda^7}\over{19372953600 }}-{{562945732229\,i\,\pi^{15}\,\lambda^6}\over{201801600}}+{{60332915\, \pi^{14}\,\lambda^5}\over{38016}} \nonumber \\
& & +{{3952891\,i\,\pi^{13}\,\lambda^4}\over{4608}} -{{4899791\,\pi^{12}\,\lambda^3}\over{11520}}-{{115583\,i\,\pi^{11}\,\lambda^2 }\over{640}}+{{7293\,\pi^{10}\,\lambda}\over{128}}+{{2431\,i\,\pi^9}\over{256 }}, \\
\langle \hat{x}^{20} \rangle_{10} &=& {{6470504221105366364567\,\pi^{20}\,\lambda^{10}}\over{96035605585920000 }}+{{611939373233888527\,i\,\pi^{19}\,\lambda^9}\over{15205637551104}}-{{337680661534598713 \,\pi^{18}\,\lambda^8}\over{14227497123840}} \nonumber \\
& & -{{1248183306883\,i\,\pi^{17} \,\lambda^7}\over{90810720}}+{{60061822300489\,\pi^{16}\,\lambda^6}\over{7749181440 }}+{{558837481\,i\,\pi^{15}\,\lambda^5}\over{133056}} \nonumber \\
& & -{{517647229\,\pi^{14 }\,\lambda^4}\over{241920}}-{{994517\,i\,\pi^{13}\,\lambda^3}\over{1008}}+{{1767779 \,\pi^{12}\,\lambda^2}\over{4608}}+{{20995\,i\,\pi^{11}\,\lambda}\over{192}}-{{4199 \,\pi^{10}}\over{256}}, \\
\end{eqnarray}

The Wilson loop is therefore 
\begin{eqnarray}
W(\lambda) &=& -{{1166161\,i\,\pi^{11}\,\lambda^{11}}\over{3628800}}+{{3447391 \,\pi^{10}\,\lambda^{10}}\over{7983360}}+{{2603\,i\,\pi^9\,\lambda^9}\over{15120 }}-{{87449\,\pi^8\,\lambda^8}\over{362880}}-{{i\,\pi^7\,\lambda^7}\over{10}}+{{151 \,\pi^6\,\lambda^6}\over{1008}} \nonumber \\
& &+{{i\,\pi^5\,\lambda^5}\over{12}}-{{29\,\pi^4\,\lambda^4 }\over{120}}-{{i\,\pi^3\,\lambda^3}\over{2}}+{{5\,\pi^2\,\lambda^2}\over{6}}+1. 
\end{eqnarray}

\end{document}